\documentclass[manuscript]{aastex63}
\usepackage{natbib}
\usepackage{fontenc}
\usepackage{comment}
\usepackage{amsmath}
\usepackage{enumitem}
\usepackage{float}
\usepackage{color}
\graphicspath{{./}{Figures/}}

\newcommand{\ree}{\textcolor[rgb]{0,0,0}}
\newcommand{\rd}{\textcolor[rgb]{0,0,0}}

\newcommand{\bl}{\textcolor[rgb]{0,0,0}}

\def\gtsima{$\;\buildrel > \over \sim \;$}
\def\simgt{\lower.5ex \hbox{\gtsima}}
\def\ltsima{$\;\buildrel < \over \sim \;$}
\def\simlt{\lower.5ex \hbox{\ltsima}}
\def \CHp{\ifmmode{\rm CH^+}\else{$\rm CH^+$}\fi}
\def \HH{\ifmmode{\rm H_2}\else{$\rm H_2$}\fi}
\def \Cp{\ifmmode{\rm C^+}\else{$\rm C^+$}\fi} 
\def \cc    {\ifmmode{\,{\rm cm}^{-3}}\else{$\,{\rm cm}^{-3}$}\fi}
\def \dens{\ifmmode{n_{\rm H}}\else{$n_{\rm H}$}\fi}
\def \kms   {\ifmmode{\,{\rm km}\,{\rm s}^{-1}}\else{km s$^{-1}$}\fi} 

\begin{document}

\title{JWST observations of cosmic-ray-excited H$_2$ in Barnard 68:\\ spatial 
variations and constraints on cosmic-ray attenuation}

\author{David A. Neufeld}
\affiliation{William H.\ Miller Department of Physics \& Astronomy, Johns Hopkins 
University, Baltimore, MD 21218, USA}

\author{Kedron Silsbee}
\affiliation{Physics Department, University of Texas at El Paso, El Paso 79968, USA}

\author{Alexei V. Ivlev}
\affiliation{Max-Planck-Institut f\"ur extraterrestrische Physik, Giessenbachstrasse 1, 85748 Garching, Germany}

\author{Shmuel Bialy}
\affiliation{Technion - Israel Institute of Technology, Haifa, 3200003
Israel}

\author{Brandt A. L. Gaches}
\affiliation{Faculty of Physics, University of Duisburg-Essen, Lotharstraße 1, 47057 Duisburg, Germany}

\author{Marco Padovani}
\affiliation{INAF-Osservatorio Astrofisico di Arcetri, Largo E. Fermi 5, 50125 Firenze, Italy}

\author{Sirio Belli}
\affiliation{Dipartimento di Fisica e Astronomia, Universit\`a di Bologna, 40125 Bologna, Italy}

\author{Thomas G. Bisbas}
\affiliation{Research Center for Astronomical Computing, Zhejiang Laboratory, Hangzhou, 311000, China}

\author{Amit Chemke}
\affiliation{Technion - Israel Institute of Technology, Haifa, 3200003
Israel}

\author{Benjamin Godard}
\affiliation{Observatoire de Paris, Universit\'e PSL, Sorbonne Universit\'e, LERMA, 75014 Paris, France}
\affiliation{Laboratoire de Physique de l’Ecole Normale Sup\'erieure, ENS, Universit\'e PSL, CNRS, 
Sorbonne Universit\'e, Universit\'e de Paris, 75005 Paris, France}

\author{James Muzerolle Page}
\affiliation{Space Telescope Science Institute, Baltimore, MD 21218, USA}

\author{Christian Rab}
\affiliation{Max-Planck-Institut f\"ur extraterrestrische Physik, Giessenbachstrasse 1, 85748 Garching, Germany}
\affiliation{University Observatory, Faculty of Physics, Ludwig-Maximilians-Universit\"at M\"unchen, Scheinerstr. 1, 81679 Munich, Germany}

\begin{abstract}
We present James Webb Space Telescope (JWST) NIRSpec observations of the 
starless dark cloud Barnard~68 that reveal the spatially-resolved 
signature of cosmic-ray excited molecular hydrogen (CRXH$_2$) emissions for the first time. 
Following up on our initial detection of CRXH$_2$ emissions from B68 (Bialy et al.\ 2025), 
we now exploit JWST’s sensitivity 
and spatial multiplexing to map CRXH$_2$ rovibrational lines across 16 sight 
lines through the cloud. By disentangling the CRXH$_2$ and UV-pumped H$_2$ components, 
we isolate the para-H$_2$--dominated spectrum attributable to cosmic-ray excitation.
We find that there are significant spatial variations in the ratio of the
CRXH$_2$ line intensity to the line-of-sight H$_2$ column density; these
cannot be accounted for by dust extinction alone
and demonstrate a clear attenuation of the cosmic-ray flux
with increasing shielding column.  Modeling B68 as a 
Bonnor--Ebert sphere, we constrain both the 
unshielded cosmic-ray ionization rate, $\zeta_{\rm H_2}$,
and how it decreases with shielding column. At a 
reference depth of $N({\rm H}_2) = 3 \times 10^{21}$~cm$^{-2}$, we infer 
$\zeta_{\rm H_2} \approx 1.4 \times 10^{-16}$~s$^{-1}$, a factor of $\approx 3$ higher 
than \rd{the average} value derived from H$_3^+$ absorption studies. 
These results provide the most direct probe to date of cosmic-ray penetration 
into cold, dense gas, offering new constraints on both the microphysics of CR--H$_2$ 
interactions and the attenuation of low-energy cosmic rays in molecular clouds. 
Our findings establish CRXH$_2$ emission as a powerful new diagnostic of the 
cosmic-ray environment in interstellar space.

\end{abstract}


\section{Introduction}

Recent JWST observations of the starless dark cloud Barnard 68 (B68)
have led to the first detection of cosmic-ray-excited H$_2$ (CRXH$_2$) 
emissions from interstellar space (S.\ Bialy et al.\ 2025; hereafter B25), 
thereby opening a new window on  Galactic cosmic rays (CR).   Because 
CRXH$_2$ emissions originate within cold ($\sim 15$~K) gas in the cloud interior, 
they exhibit a spectrum that is readily distinguished from UV excited H$_2$ (UVXH$_2$) 
emissions from warmer ($\sim 100$~K) gas at the cloud surface: CRXH$_2$ exhibits a spectrum that is
strongly dominated by para-H$_2$ (even-$J$) lines (S.\ Bialy et al. 2020, hereafter
B20; M. Padovani et al. 2022, hereafter P22) whereas UVXH$_2$ emits lines of both ortho- 
and para-H$_2$ with an ortho-to-para ratio (OPR) $\approx 1.6$ 
(e.g.\ Sternberg 1988, hereafter S88).  The spectrum of B68 measured with the NIRSpec
multiobject spectrograph (MOS) on JWST showed (B25) a clear excess of CRXH$_2$ lines, implying
an average cosmic-ray ionization rate, $\zeta_{{\rm H}_2}$, of $1.7 \times 10^{-16}\,\rm 
s^{-1}$.  This value is the total rate of ionization of H$_2$, including secondary ionizations.
\rd{B25 also observed a nearby “background” region, where the column density is much lower than
at the B68 position. Because the CR signal scales with the column density (B20),
we expect the background region to be dominated by UVXH$_2$ emissions and indeed
the relative H$_2$ line intensities measured at the background position were 
fully consistent with the UV model predictions.}

In the present study, we have extended our analysis to include spatial variations in the
CRXH$_2$ emissions.  The exquisite sensitivity of JWST/NIRSpec is sufficient to permit 
high signal-to-noise ratio (SNR) detections of H$_2$ lines at multiple positions
across B68, allowing variations in $\zeta_{{\rm H_2}}$ to be inferred. These, in turn, 
provide important constraints on the propagation of CR from the diffuse outer regions
of the B68 cloud to the dense interior.

In Section 2, we describe the data acquisition, reduction and analysis methods that were 
adopted.  In Section 3, we present the line intensities measured as a function of
position on the sky.  In Section 4, we compare the observed spatial variations in the line 
intensities with a parameterized model for the attenuation of CR
that also includes the effects of dust extinction.  A discussion follows in Section 5,
and the conclusions of our study are summarized in Section 6.

\section{Observations and data reduction}

As discussed in B25, all the data were acquired in JWST program 5064, using
the NIRSpec MOS with the G235H/F170LP grating/filter combination.  The latter
provides almost complete coverage of the 1.66 - 3.05 $\mu$m spectral region at
a spectral resolving power, $\lambda/\Delta \lambda$ of 2700, sufficient to resolve
individual rotational lines within the H$_2$ $v = 1-0$ Q-branch around 2.4 $\mu$m.  
The MOS allows spectra to be obtained in separate shutters that may be opened
at commandable positions within a $3.4 \times 3.6$$''$ region on the sky.  For the present
study, we arranged for 382 open shutters to lie on a linear locus across the B68 cloud. 
Because the four quadrants of the detector do not perfectly abut one another, there is a central gap
in the spatial coverage, of length 37$''$, and a small ($\sim 0.1\,\mu$m) 
gap in the wavelength coverage around 2.25$\,\mu$m.  In Figure 1, we show the locus of 
the shutters (white lines) superposed on an extinction map of the B68 cloud 
(J. Alves et al. 2025, private communication).  In a separate observation, we used the
same arrangement of shutters to observe a background region (hereafter the ``OFF" position)
located 30$'$ due north of B68.   

\begin{figure}
\centering
\includegraphics[width=0.8\linewidth]{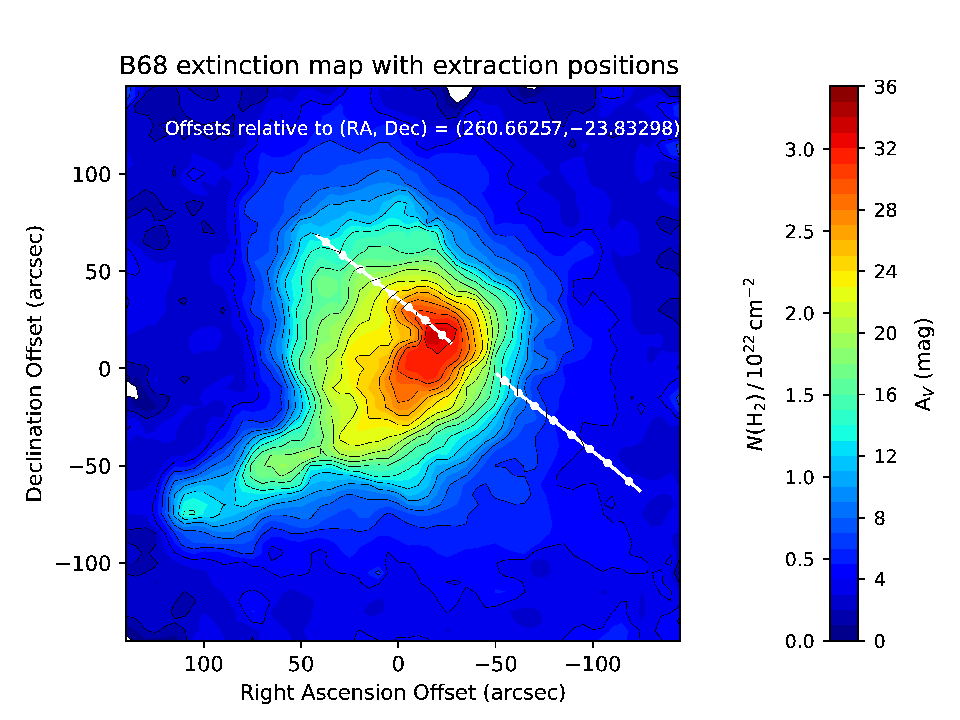}
\caption{Arrangement of NIRSpec MOS shutters across Barnard~68. White lines show 
the locus of the 382 open shutters used to obtain spectra across the cloud. 
The 16 averaged slitlets ($\sim$11$''$ each) are marked with white dots. 
The background (``OFF'') position, observed 30$'$ north of B68, used the same configuration. 
The extinction map is adapted from Alves et al. (2025, private comm.).}
\end{figure}        

As discussed in B25,
all the observations were conducted using the NRSIRS2 readout mode, with 4 
integrations of 20 groups each at five dither positions on B68, providing a total 
on-source time of 8.2 hours. The dither pattern shifted the pointing by five shutters 
along both spatial axes to reduce contamination from nearby stars and to 
average over detector defects.  Measurements of the background region
employed three dither positions, 
each consisting of a single 20-group integration, for a total exposure time of 1.2 hours.

All data were reprocessed beginning with the raw ``uncal" files using the JWST calibration 
pipeline (v1.16.1) together with CRDS context 1303. Standard settings were applied 
throughout, except during outlier rejection, where weighting by ``exptime" 
substantially reduced the number of residual outliers relative to the default configuration. 
The MSA metadata were adjusted so that the pipeline could align and extract spectra from 
each open shutter of the long slit across all dither positions in both target and 
background fields. 

Average spectra were extracted separately for each of 16 slitlets of length $\sim$ 11$''$, 
each formed from $\sim 24$ 
adjacent shutters on B68; the central positions for these 16 slitlets
are marked by white dots.   We also obtained an average spectrum for the entire ensemble of
slitlets, both for slitlets on B68 and for those on the OFF-position.  The difference 
spectrum, ``B68 -- OFF," was presented in B25 (their Figure 1b).  A total of 21 spectral 
lines were securely detected in the ensemble average spectrum for B68.  These comprise 
(1) eleven lines in the $v = 1-0$ band of H$_2$ -- one of which, $v = 1-0 S(0)$, lies in the
detector gap near 2.25~$\mu$m and can be observed only in a subset of the slitlets and was 
therefore excluded from the analysis; (2) five lines in the $v = 2-1$ band of H$_2$; 
(3) four lines in the $v = 3-2$ band of H$_2$; (4) the Paschen alpha ($n=4-3$)
line of atomic hydrogen.  

\section{Results}
\subsection{Average H$_2$ line intensities}

We obtained Gaussian fits to each of these lines, using the 
Levenberg-Marquardt algorithm to fit a flat baseline plus a Gaussian to each line. 
Here,  there were 4 free parameters: the line width (FWHM in $\rm km\,s^{-1}$), the integrated line 
intensity ($\rm erg\,cm^{-2}\,s^{-1}\,sr^{-1}$), the
line centroid ($\rm km\,s^{-1}$ in the barycentric frame), and the continuum level  
(MJy~sr$^{-1}$).  Each fit was optimized over the barycentric velocity interval 
$[-600, 600]\rm \, km\,s^{-1}.$   \rd{In computing the fits to the $v=1-0$~O(2) line, we corrected for contamination by
the partially blended H Br-$\beta$ line, which has a rest wavelength that is only 
$1.015\times 10^{-3}\,\mu$m smaller 
than that of $v=1-0$~O(2), corresponding to a blueshift of $116\,\rm km\,s^{-1}$. 
This correction was made by assuming a Br-$\beta$ 
line width and a barycentric centroid velocity identical to that observed for H Pa-$\alpha$ and
a Br-$\beta$/Pa-$\alpha$ flux ratio of 0.134, the value expected for Case B recombination 
(Storey and Hummer 1995)}

Figure 2 shows the line profiles (blue histogram) and fitted 
Gaussians (red) for  9 example lines, and Table 1 lists the fit parameters with 1$\sigma$ 
statistical errors. 
Figure 3 shows the line FWHM (km/s) and centroid velocities (km/s in the barycentric
frame) for the Gaussian fits shown in Figure 2 as a function of line wavelength. Here, para-H$_2$ lines appear in red and ortho-H$_2$ lines appear in blue. The dashed black line in the upper
panel is a quadratic fit to the line FWHM. It is in good agreement with what is expected for
an unresolved spectral line (Jakobsen et al. 2022, their Fig. 5). The horizontal
lines in the lower panel show the mean velocity centroids ($-15.9$, $-10.2$, $-14.4 \, \rm km\,s^{-1}$,
respectively), for the ortho-H$_2$ lines (blue), para-H$_2$ lines (red), and for all H$_2$ lines (black).
There is no significant difference between the average centroid velocities for the ortho- and
para-lines, and the line-to-line variation of velocity centroids is consistent with the
expected systematic calibration errors\footnote{https://jwst-docs.stsci.edu/jwst-calibration-status/nirspec-calibration-status, downloaded 2025 May 2}
of 15 km/s.  \rd{Within the expected errors, the velocity centroids also agree with
the systemic velocity of B68 determined by Alves et al.\ (2001; hereafter A01): 
$\rm +3.4\, km\,s^{-1}$ relative to the local standard of rest, corresponding to a barycentric velocity of 
$\rm -8.2\, km\,s^{-1}$.}

Because the results shown in Figure 3 support our strong expectations that (1) the
lines are spectrally unresolved with NIRSpec, and (2) the velocity centroids are
the same for all lines, we decided to adopt these expectations as {\it priors} in the 
fitting process and subsequently refit all lines with the velocity centroid fixed at
$\rm -14.4\, km\,s^{-1}$ and the line width constrained to the (wavelength-dependent) values shown by the
dashed line in the upper panel of Figure 3.  
Only the continuum level and line intensity
are now allowed to vary in the optimization of the fit.  This approach reduces the uncertainties
in the line intensities and is particularly advantageous for weak lines.  { For strong lines, 
no significant effects on the inferred line parameters could be discerned.}
The resultant line
intensities are tabulated in Table 2 for the B68, OFF and B68 -- OFF spectra.  

\subsection{Separation of the UVXH$_2$ and CRXH$_2$ emissions}

Figure 4 (upper panel)
shows the integrated line intensities for the OFF position, $I({\rm OFF}) = 10^{-7} I_{-7}({\rm OFF})
\, \rm erg\,cm^{-2}\,s^{-1}\,sr^{-1}$, each divided by 
$f_{\rm UV}$, defined here as the fraction of the rovibrational UVXH$_2$
line emission that is predicted to emerge in each line.  Blue and red points are
for transitions of ortho- and para-H$_2$ respectively.
Here, we used the values for $f_{\rm UV}$ computed by S88 for a 
gas temperature of 100~K, and the 
blue and red horizontal lines
show the mean values of $I_{-7}({\rm OFF})/f_{\rm UV}$ for ortho- and para-H$_2$.  Figure 4 indicates that the 
OFF position line strengths are in good agreement with the predictions for pure 
UV fluorescence in gas with a temperature 100~K (in which the ortho-to-para ratio
is $\approx 1.6$).  Our estimate for the summed intensity for all H$_2$ rotational lines 
(both observed and unobserved) is $(30 \pm 2) \times 10^{-7}\,
\rm erg\,cm^{-2}\,s^{-1}\,sr^{-1}.$  

Because CRXH$_2$ emissions originate
primarily in the cold cloud interior where the OPR is $\ll 1$, they make a
negligible contribution to the ortho-H$_2$ line strengths. 
For the difference spectrum, B68 -- OFF, the ortho-H$_2$ lines are weaker
than in either the B68 or OFF spectra but the cancellation is incomplete; this
presumably reflects differences in the
incident UV field and/or the geometry (e.g. the degree of limb brightening).
The overall ortho-H$_2$ line intensities in the difference spectrum are smaller than those
in the \bl{OFF} spectrum: on average, the ratio, $Q$, of the ortho-H$_2$ line intensities 
in the difference spectrum to those in the \bl{OFF} spectrum is $Q = 0.445$.  We may therefore 
estimate the CRXH$_2$ line emissions from B68 by subtracting 0.445 times the OFF line
intensity from the intensity observed in the \bl{difference spectrum}.  { Here, we
assume that the UVXH$_2$ line ratios are the same at the two observed
positions, B68 and OFF.}
The resultant intensities,
attributable to CR excitation alone, 
are shown in the bottom panel of Figure 4.  
Unequivocal detections 
(with a signal-to-noise ratio, SNR, greater than 5) are obtained for CRXH$_2$ emission in
the $v = 1-0$ O(2), O(4), Q(2) and S(2) lines.  CRXH$_2$ emissions are also
securely detected in the $v = 1-0$ Q(4) line and are tentatively detected ($\rm 4 < SNR < 5$) in
the $v = 2-1$ O(2) and Q(2) lines.  The strongest line, $v = 1-0$ O(2),
has a CR contribution that is detected in the difference spectrum at the 28$\,\sigma$ level.

\subsection{Spatial variation}

To probe the spatial variation of the CRXH$_2$ emissions, 
we obtained line intensities for each of the 16 individual slitlets,
corresponding to 16 positions along the slit (Figure 1).  Here, we adopted the method
used above to compute the intensities plotted in the lower panel of Figure 4, but we now
implemented it separately for each slitlet (with account taken for spatial variations
in the value of $Q$).  In Figure 5, results are shown for the
three strongest CRXH$_2$ lines. The intensities attributable to CRXH$_2$, after subtraction of the
UVXH$_2$ contribution, are shown as a function of the position along the slit. 

\begin{figure}
\centering
\includegraphics[width=0.9\linewidth]{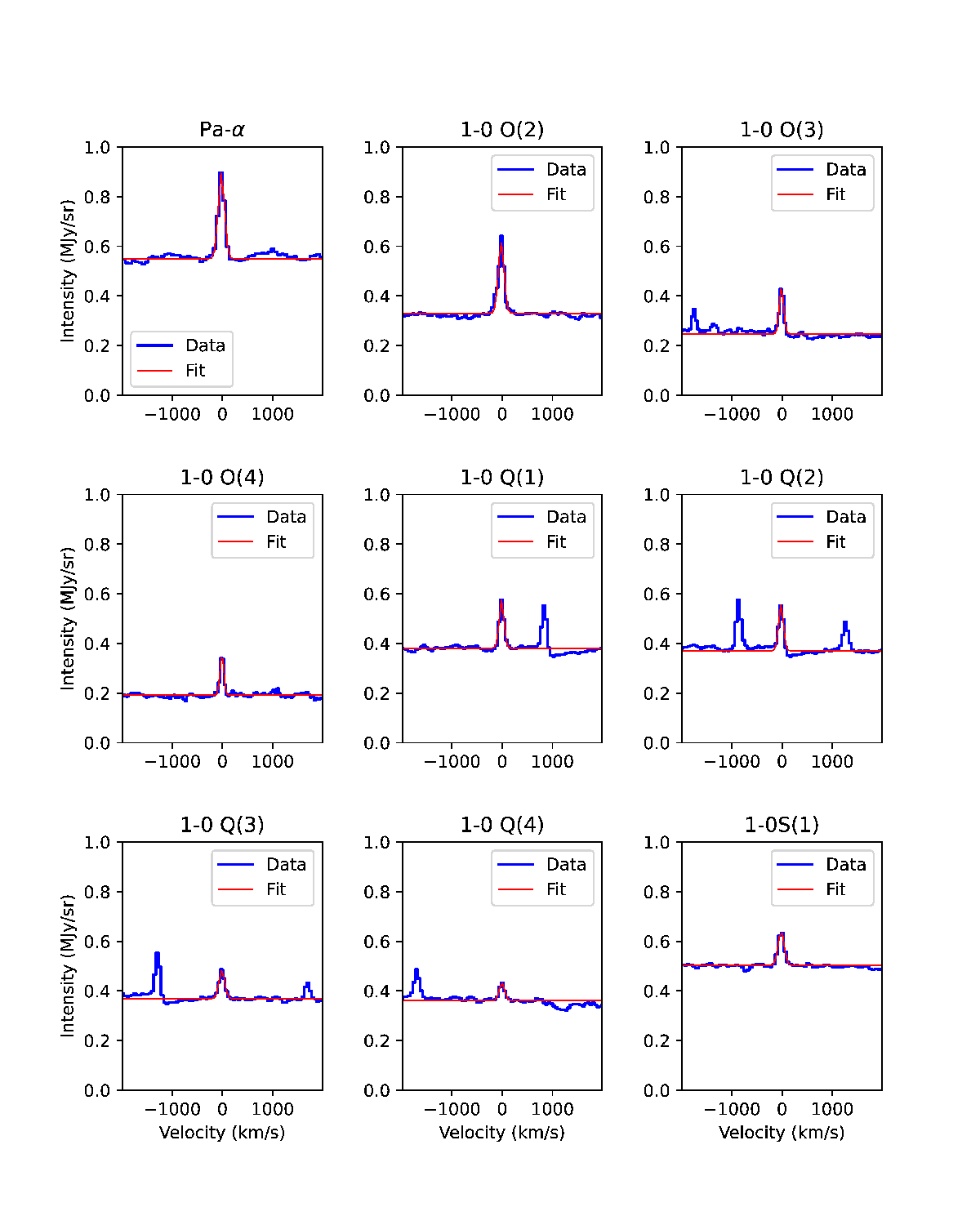}
\caption{Example H$_2$ rovibrational line profiles from B68 (black histograms) 
with Gaussian fits (red curves).}
\end{figure}

\begin{table*}[htbp]
\centering
\caption{Gaussian fit results for detected H$_2$ and H\,{\sc i} lines in B68.
Listed are the transition, rest wavelength, integrated line intensity, 
velocity centroid, and FWHM.}
\label{tab:linefits_vel_sorted}
\begin{tabular}{lccrr}
\hline
Transition & $\lambda_{\rm rest}$ & Intensity & $v_{\rm cent}$\phantom {00} & FWHM \\
 & ($\mu$m) & ($10^{-7}$ erg cm$^{-2}$ s$^{-1}$ sr$^{-1}$) & (km\,s$^{-1}$) & (km\,s$^{-1}$) \\
\hline
Pa-$\alpha$     & 1.87561 & $2.90 \pm 0.076$ & $-20 \pm 2$ & $147 \pm 4 \phantom{0}$ \\
1--0 S(3)       & 1.95756 & $0.59 \pm 0.040$ & $-28 \pm 5$ & $168 \pm 12$ \\
1--0 S(2)       & 2.03376 & $0.59 \pm 0.056$ & $-21 \pm 5$ & $134 \pm 13$ \\
1--0 S(1)       & 2.12183 & $0.97 \pm 0.029$ & $-18 \pm 2$ & $139 \pm 4 \phantom{0}$ \\
1--0 Q(1)       & 2.40659 & $0.95 \pm 0.036$ & $-12 \pm 2$ & $111 \pm 4 \phantom{0}$ \\
1--0 Q(2)       & 2.41344 & $0.92 \pm 0.082$ & $-21 \pm 4$ & $112 \pm 11$ \\
1--0 Q(3)       & 2.42373 & $0.66 \pm 0.030$ & $-11 \pm 3$ & $133 \pm 6 \phantom{0}$ \\
1--0 Q(4)       & 2.43749 & $0.39 \pm 0.029$ & $-6 \pm 4$  & $125 \pm 10$ \\
3--2 S(0)       & 2.50144 & $0.33 \pm 0.022$ & $9 \pm 4$   & $143 \pm 10$ \\
2--1 Q(1)       & 2.55099 & $0.49 \pm 0.020$ & $-25 \pm 2$ & $133 \pm 6 \phantom{0}$ \\
2--1 Q(2)       & 2.55851 & $0.43 \pm 0.021$ & $-12 \pm 3$ & $127 \pm 6 \phantom{0}$ \\
2--1 Q(3)       & 2.56983 & $0.20 \pm 0.020$ & $-10 \pm 5$ & $99 \pm 11$ \\
1--0 O(2)       & 2.62688 & $1.55 \pm 0.095$ & $-20 \pm 4$ & $135 \pm 9 \phantom{0}$ \\
3--2 Q(1)       & 2.71025 & $0.19 \pm 0.038$ & $-18 \pm 9$ & $104 \pm 23$ \\
3--2 Q(2)       & 2.71862 & $0.15 \pm 0.015$ & $-11 \pm 4$ & $90 \pm 10$ \\
2--1 O(2)       & 2.78616 & $0.29 \pm 0.032$ & $-5 \pm 4$  & $87 \pm 10$ \\
1--0 O(3)       & 2.80252 & $0.73 \pm 0.045$ & $-12 \pm 3$ & $102 \pm 7\phantom{0}$ \\
3--2 O(2)       & 2.96206 & $0.14 \pm 0.018$ & $-8 \pm 5$  & $81 \pm 11$ \\
2--1 O(3)       & 2.97406 & $0.28 \pm 0.023$ & $-11 \pm 4$ & $99 \pm 9\phantom{0}$ \\
1--0 O(4)       & 3.00387 & $0.53 \pm 0.021$ & $-10 \pm 2$ & $87 \pm 4\phantom{0}$ \\
\hline
\end{tabular}
\end{table*}

\begin{figure}
\centering
\includegraphics[width=0.8\linewidth]{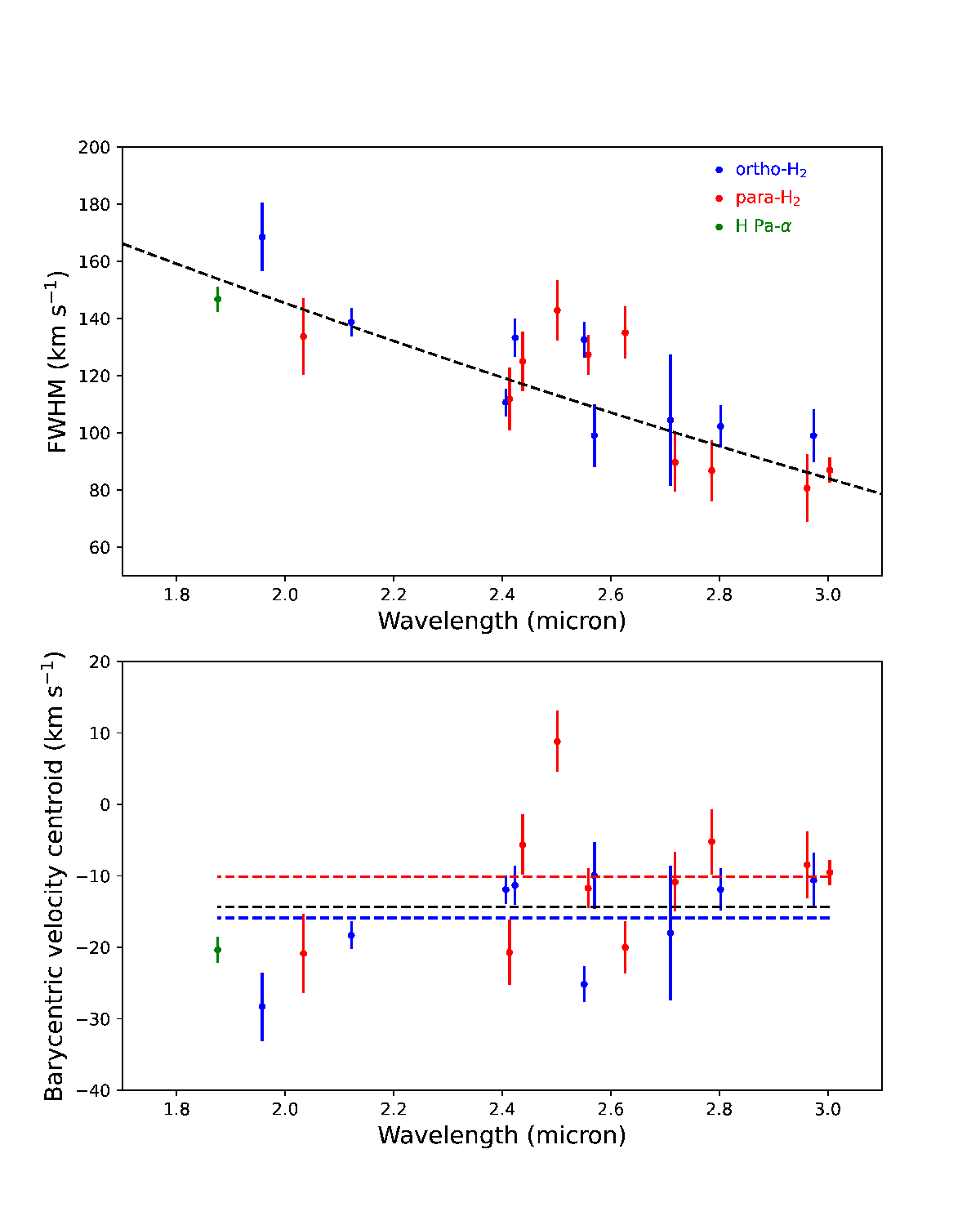}
\caption{Measured line widths (upper panel) and centroid velocities (lower panel) as a 
function of wavelength for all detected H$_2$ lines. Para-H$_2$ lines are shown in red, 
and ortho-H$_2$ lines in blue. The dashed curve in the upper panel is a quadratic 
fit to the unresolved instrumental profile. In the lower panel, horizontal lines 
show mean velocity centroids for ortho-, para-, and all H$_2$ lines; the 
scatter is consistent with calibration uncertainties.}
\end{figure}

\begin{table*}
\centering
\caption{Average Line Intensities ($10^{-7}\rm\,erg\,cm^{-2}\,s^{-1}\,sr^{-1}$) with line kinematics fixed  }
\begin{tabular}{lcccc}
\hline
Transition & $\lambda$ ($\mu$m) & B68 & OFF &  B68 -- OFF \\
\hline
Pa-$\alpha$ & 1.87561 & $2.976 \pm 0.083$ & $1.905 \pm 0.138$ & $1.071 \pm 0.103$ \\
1--0 S(3)   & 1.95756 & $0.534 \pm 0.035$ & $0.401 \pm 0.037$ & $0.133 \pm 0.031$ \\
1--0 S(2)   & 2.03376 & $0.615 \pm 0.044$ & $0.226 \pm 0.026$ & $0.389 \pm 0.047$ \\
1--0 S(1)   & 2.12183 & $0.963 \pm 0.024$ & $0.623 \pm 0.038$ & $0.340 \pm 0.038$ \\
1--0 Q(1)   & 2.40659 & $0.993 \pm 0.031$ & $0.600 \pm 0.040$ & $0.393 \pm 0.034$ \\
1--0 Q(2)   & 2.41344 & $0.943 \pm 0.066$ & $0.293 \pm 0.058$ & $0.650 \pm 0.028$ \\
1--0 Q(3)   & 2.42373 & $0.613 \pm 0.023$ & $0.448 \pm 0.049$ & $0.166 \pm 0.043$ \\
1--0 Q(4)   & 2.43749 & $0.375 \pm 0.023$ & $0.151 \pm 0.023$ & $0.224 \pm 0.027$ \\
3--2 S(0)   & 2.50144 & $0.269 \pm 0.025$ & $0.143 \pm 0.029$ & $0.127 \pm 0.025$ \\
2--1 Q(1)   & 2.55099 & $0.430 \pm 0.022$ & $0.300 \pm 0.035$ & $0.130 \pm 0.031$ \\
2--1 Q(2)   & 2.55851 & $0.392 \pm 0.016$ & $0.168 \pm 0.031$ & $0.224 \pm 0.031$ \\
2--1 Q(3)   & 2.56983 & $0.214 \pm 0.017$ & $0.132 \pm 0.028$ & $0.083 \pm 0.028$ \\
1--0 O(2)   & 2.62688 & $1.280 \pm 0.036$ & $0.290 \pm 0.034$ & $0.990 \pm 0.027$ \\
3--2 Q(1)   & 2.71025 & $0.184 \pm 0.028$ & $0.169 \pm 0.034$ & $0.015 \pm 0.027$ \\
3--2 Q(2)   & 2.71862 & $0.161 \pm 0.012$ & $0.100 \pm 0.023$ & $0.061 \pm 0.024$ \\
2--1 O(2)   & 2.78616 & $0.302 \pm 0.028$ & $0.107 \pm 0.033$ & $0.195 \pm 0.028$ \\
1--0 O(3)   & 2.80252 & $0.701 \pm 0.034$ & $0.545 \pm 0.033$ & $0.156 \pm 0.031$ \\
3--2 O(2)   & 2.96206 & $0.146 \pm 0.014$ & $0.155 \pm 0.020$ & $-0.010 \pm 0.019$ \\
2--1 O(3)   & 2.97406 & $0.254 \pm 0.017$ & $0.215 \pm 0.037$ & $0.039 \pm 0.035$ \\
1--0 O(4)   & 3.00387 & $0.512 \pm 0.019$ & $0.208 \pm 0.032$ & $0.304 \pm 0.036$ \\
\hline
\end{tabular}
\end{table*}

\begin{figure}
\centering
\includegraphics[width=0.8\linewidth]{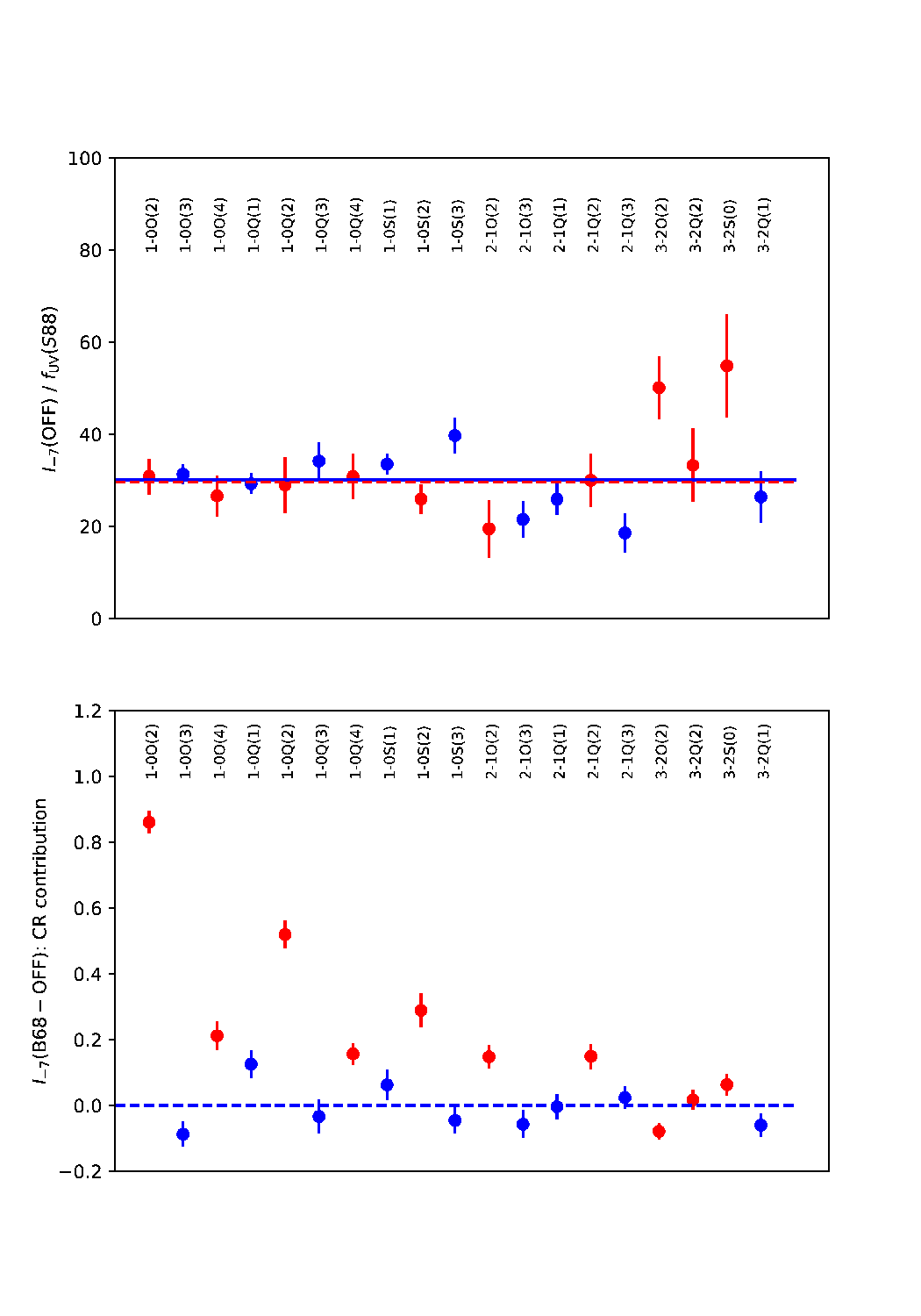}
\caption{Separation of UV-excited and cosmic-ray--excited H$_2$ emission. 
Top: integrated line intensities at the OFF position, divided by predicted UV 
fluorescence fractions, from the 100~K models of Sternberg (1988).  Ortho-H$_2$ and 
para-H$_2$ transitions are shown in blue and red.  Bottom panel: residual B68 line strengths 
attributable to CR excitation.}
\end{figure}

\begin{figure}
\centering
\includegraphics[width=0.8\linewidth]{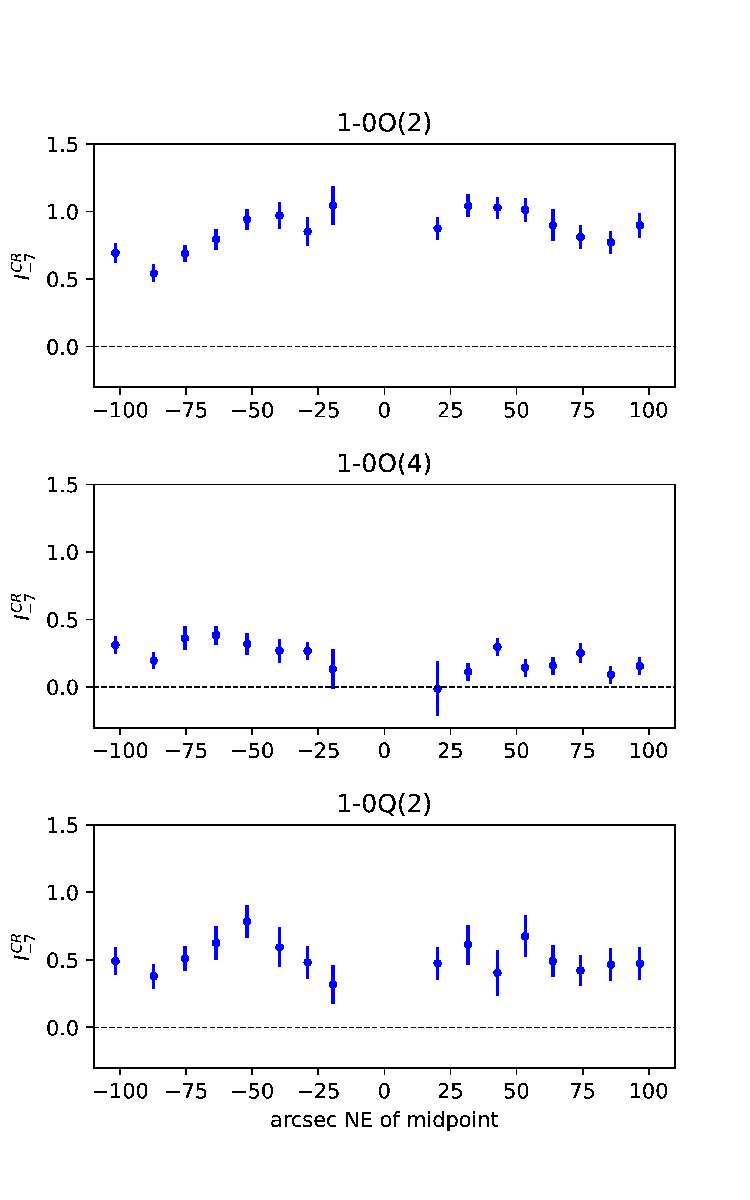}
\caption{Spatial variation of CR-excited H$_2$ line intensities across B68. 
Shown are three of the strongest lines, plotted as a function of position along the slit (see Figure~1). 
Intensities are corrected for UV contamination using the OFF position, {and are given in units
of $10^{-7}\,\rm erg\,cm^{-2}\,s^{-1}\,sr^{-1}$.}}
\end{figure}

In Figure 6, we show the ratio of these intensities to the column density of H$_2$ along the
line of sight, $$N({\rm H}_2)= 10^{22}\,N_{22} \, \rm cm^{-2}.$$
The latter was obtained from the extinction map, assuming that the
gas is fully molecular with $N({\rm H}_2) = \onehalf N_{\rm H}$ and adopting 
an $N_{22}/A_V$ ratio of 0.094\footnote{\bl{This value corresponds to 
$N_{\rm H}/A_V = 1.87 \times 10^{21}\,\rm cm^{-2}$, the canonical $N_{\rm H}/A_V$ ratio
for the {\it diffuse} ISM.
This ratio is based on the $N_{\rm H}/E(B-V)$ ratio obtained from UV absorption
measurements of H and H$_2$ (Bohlin et al.\ 1978) and an assumed 
$A_V / E(B-V)$ ratio of 3.1.  X-ray absorption observations 
(e.g. Zhu et al.\ 2017 and references therein) yield similar values for $N_{\rm H}/A_V$
in the diffuse ISM.  For the {\it dense} ISM, some studies 
have favored a smaller value of $N_{\rm H}/A_V$; guided by dust models from Draine (2003),
Evans et al.\ (2009)
assumed $N_{\rm H}/A_V = 1.37 \times 10^{21}\,\rm cm^{-2}$, about $27\%$ smaller 
than the value we adopt here.
On the other hand, direct measurements of ${\rm H}_2$ in the 
Taurus molecular cloud, obtained by observing its weak near-IR quadrupole transitions in 
absorption (Lacy et al. 2017), yielded a mean $N({\rm H}_2)/E(J-K)$ ratio of 
$5.4 \times 10^{21}\,\rm cm^{-2}$, from
which a $N({\rm H}_2)/A_V$ ratio of $1.0 \times 10^{21}\,\rm cm^{-2}$ was inferred.  This, 
then, is slightly (6$\%$) {\it larger} than the ratio we assume.}}.
\rd{The column densities determined from the extinction map are in acceptable
agreement with independent determinations based on submillimeter dust emission (Nielbock et al.\
2012; Roy et al. 2014), although the latter are also dependent on the assumed dust properties 
(i.e.\ the submillimeter opacity law).}

The decline of $I_{-7}/N_{22}$ with $N_{22}$ has two possible causes that must be disentangled: 
(1) a decline in $\zeta_{\rm H_2}$ with increasing depth into the cloud; and (2) dust
extinction of the emergent H$_2$ emissions.  The next section describes a method for
separating these two effects.

\begin{figure}
\centering
\includegraphics[width=0.8\linewidth]{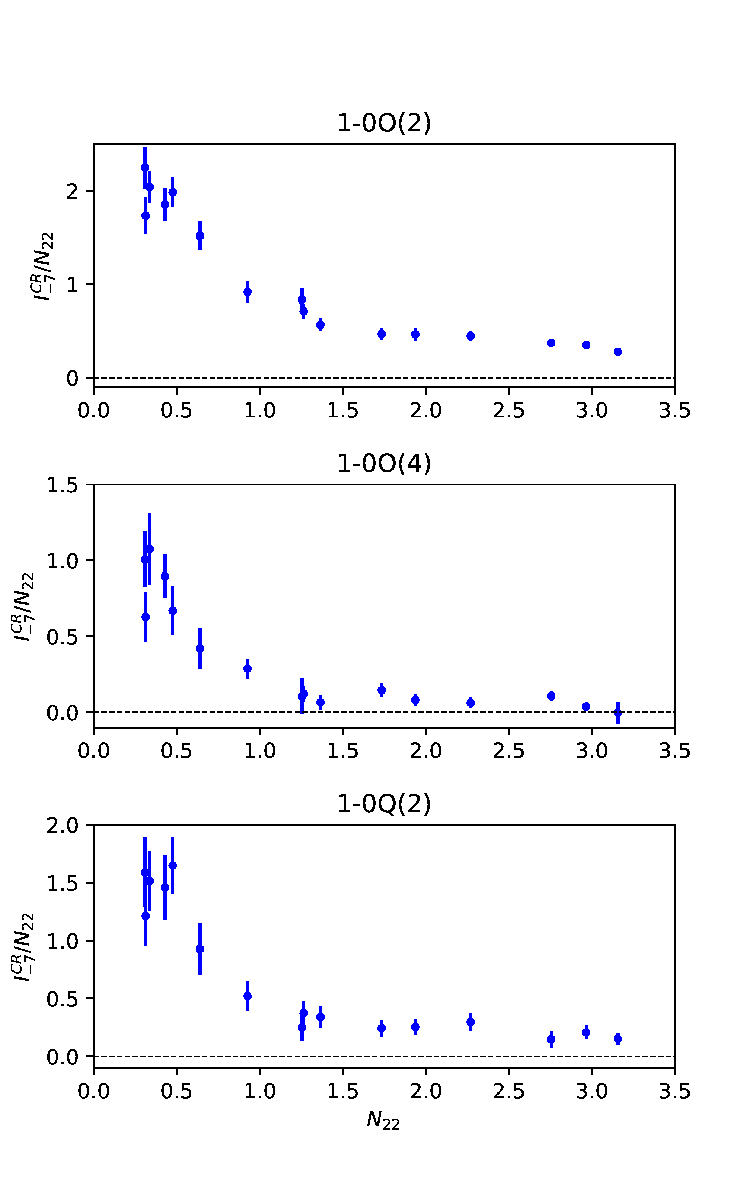}
\caption{Ratio of CR-excited H$_2$ line intensity to H$_2$ column density, 
$I_{-7}/N_{22}$, as a function of $N_{22}$. The decrease with increasing column density 
reflects the combined effects of cosmic-ray attenuation and dust extinction.}
\end{figure}

\section{Cloud model and inferred attenuation of CR}

\subsection{Cloud model}
To analyze the spatial variation of the observed CRXH$_2$ emission, we used
a simple model in which the Barnard 68 cloud was assumed to be a 
fully-molecular Bonnor-Ebert sphere.
This assumption is motivated by the pioneering study of A01, who
analysed an extinction map they obtained for B68.
They found that a Bonnor-Ebert sphere with a dimensionless outer radius 
$\xi_{\rm out}=6.9$ provided an excellent fit to the azimuthally-averaged extinction 
as a function of projected angular distance.  The dimensionless radius, $\xi$ is related
to the physical radius, $r$, the isothermal sound speed, $c_s$, and the central H$_2$
density $n_0$, by the expression
$$\xi = \frac{r}{c_s} \sqrt{4\pi G \mu n_0},$$
where $G$ is the gravitational constant and $\mu = 4.58\times 10^{-24}\,\rm g$ is the mean mass per
H$_2$ molecule (including helium).  
The best fit value for $\xi_{\rm out}$ obtained
by A01 was close to -- but somewhat larger than -- the critical value of 6.45 above 
which a (non-magnetized)
cloud becomes unstable.
The volume density, $n_{{\rm H}_2}$, was obtained as a function of
\rd{the distance from the center of the cloud}, $r$, by solving the Lane-Emden equation for an isothermal sphere in hydrostatic equilibrium.

In the following analysis, we adopt a Cartesian coordinate system with the $x$-axis
in the plane-of-the-sky along the slit, the $z$-axis along the line-of-sight with
$z$ values increasing with distance from the observer, and the $y$-axis perpendicular to 
$x$ and $z$.  The origin is at the center of the cloud, and the slit is assumed to lie
at $y=0$.  \rd{The radial coordinate, $r$, is then $(x^2+y^2+z^2)^{1/2}$.}
The total column density along the line of sight, $N_{{\rm H}_2}$,
was obtained as a function of projected distance, $x$, by integrating the volume density 
$$N_{{\rm H}_2}(x) = \int n_{{\rm H}_2}(r) dz = \int n_{{\rm H}_2}([x^2+z^2]^{1/2}) dz$$
In Figure 7, the results for $n_{{\rm H}_2}(r)$ (blue curve) and $N_{{\rm H}_2}(x)$
(red curve) are shown as a function
of $r/r_{\rm out}$ or $x/r_{\rm out}$, where $r_{\rm out}$ is the outer radius of the cloud. 

The quantities plotted in Figure 7 are dimensionless.  To completely specify the
properties of the B68 cloud, two of the following four dimensional quantities must be 
specified: the outer radius, $r_{out}$, the maximum H$_2$ column density, $N(0)$,
the central density, $n(0)$, or the sound speed, $c_s$.  The first two quantities are
closest to the observations, and we adopt 
$r_{\rm out}=1.25 \times 10^4 \rm \, a.u. = 1.87 \times 10^{17} \, cm$ and
$N(0) = 2.82 \times 10^{22} \rm cm^{-2}$, following A01.
These values imply a central density of $1.92 \times 10^5\, \rm cm^{-3}$ and an isothermal
sound speed, $c_s$, 
of $\rm 0.233\,km\, s^{-1}$; the latter corresponds to a 
gas temperature of 18~K if thermal pressure is dominant.\footnote{This value 
significantly exceeds temperature estimate of 10~K obtained by Hotzel et al.\ (2002)
from NH$_3$ line ratios.  This discrepancy might be resolved if the source distance
is considerably smaller than that assumed by A01 or if non-thermal support contributes
significantly: while the NH$_3$ linewidths measured by Hotzel et al.\ rule out turbulence
as a significant source of non-thermal support, magnetic fields may be 
important here (Kandori et al.\ 2020)}

\begin{figure}
\centering
\includegraphics[width=0.8\linewidth]{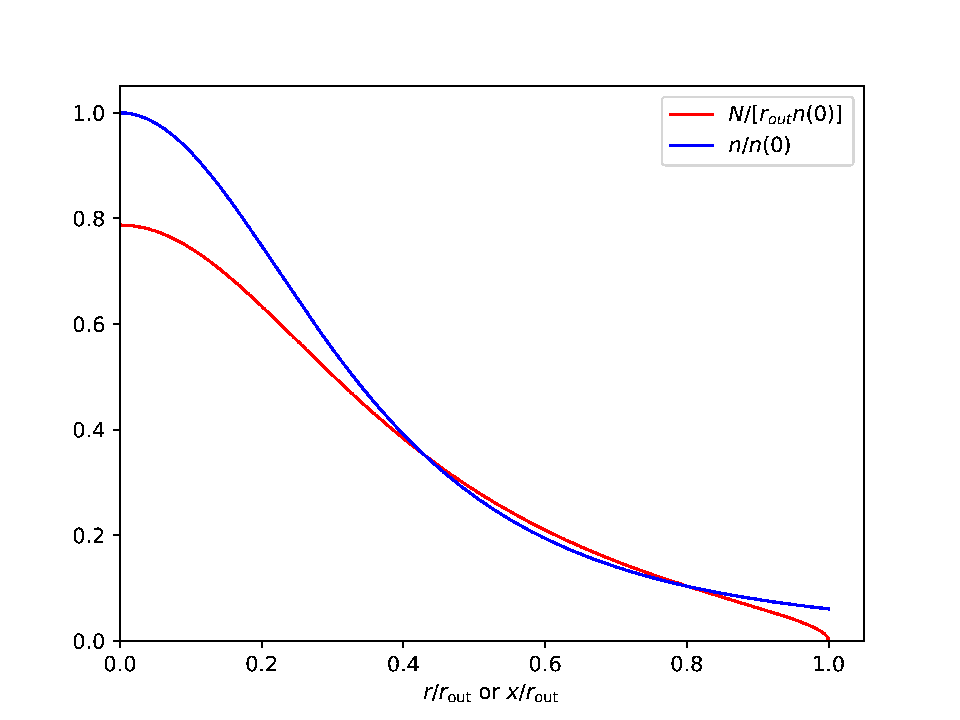}
\caption{Structure of the Bonnor--Ebert sphere model used to describe B68.  Blue curve: radial 
density profile, normalized to the central density. Red curve: normalized column density 
as a function of projected radius. The adopted model is consistent with the extinction 
profile of Alves et al. (2001).}
\end{figure} 

The H$_2$ emission from such a sphere was computed, as a function of $x$, with the inclusion 
of CR attenuation and dust extinction.   We treated the effects of CR attenuation by adopting a simple parameterization for the
reduction in the cosmic ray ionization rate due to CR propagating through a shielding 
H$_2$ column density of $N_s$:
$$\zeta_{{\rm H}_2}={\zeta_0 \over (1 + N_s/N_0)^\alpha} = \zeta_0 F_s(N_s),$$
where $\zeta_0$, $N_0$ and $\alpha$ are parameters that will be adjusted to optimize
the fit to the data.  \ree{We note here that the shielding column, $N_s$, is different from
the column density along a given sightline; the former varies with position along the
sightline and the CR propagation direction.}
\bl{The power-law index $\alpha$ describes the rate at which $\zeta_{{\rm H}_2}$ decreases
with shielding column in the limit of large $N_s$, while the parameter $N_0$ is the
shielding column below which $\zeta_{{\rm H}_2}$ is constant.  If the intrinsic 
energy spectrum of the CR at the site of their production is a power-law extending to
low energies, $N_0$ is naturally interpreted as the column density through which the
CR have passed in propagating from their production site to the surface of the B68
molecular cloud.}

For the transition from state $u$ to $l$, 
the emergent intensity of CRXH$_2$ at projected distance $x < r_{\rm out}$ 
from the cloud center is 
$$I_{ul}(x) = {b_{u} \alpha_{ul} 
hc \over 4 \pi \lambda_{ul}} \int_{-z_0}^{+z_0} \zeta_{{\rm H}_2})(x,z^\prime) 
n_{{\rm H}_2}([x^2+z^{\prime 2}]^{1/2}) e^{-\tau_d} dz^\prime$$ 
where $z$ is the distance along the line-of-sight from the cloud midplane;
$z_0=(r_{\rm out}^2 - x^2)^{1/2}$ is the half thickness of the cloud at projected distance, $x$;
$$\tau_d(x,z) =  \sigma_d \int_{-z_0}^{z} n_{{\rm H}_2}  dz^\prime$$
is the extinction optical depth from the cloud surface to position $z$ along the line-of-sight;
and $\sigma_d$ is the dust extinction cross-section per H$_2$ molecule at the wavelength, $\lambda_{ul}$,
of the H$_2$ line.  To relate $A_\lambda$  to the visual extinction, $A_V$, we adopted
the KP5 extinction curve tabulated in Pontoppidan et al.\ (2024): this extinction curve is considered
most appropriate to dark clouds such as B68.  For simplicity, our treatment adopts
a "foreground screen" geometry and thus does not include the diffuse scattered CRXH$_2$ 
component.  
The quantities $b_{u}$ and $\alpha_{ul}$ are specific to each transition 
(B25; see their Table S1): they are respectively
the excitation rate to state $u$ divided by $\zeta_{{\rm H}_2}$; and the probability
that radiative decay of state $u$ leads to state $l$. 

The three-dimensional dependence of $\zeta_{{\rm H}_2}$ depends on the
direction in which the CR travel.  We considered four cases.  In cases X, Y, and Z,
CR are assumed to propagate along the $x$, $y$ and $z$  axes, 
respectively, with equal fluxes in the $\pm$ directions.  In Case Q (quasi-isotropic), we assumed
equal contributions for all six directions, $\pm x$, $\pm y$ and $\pm z$.  Expressions 
for $\zeta_{{\rm H}_2}$ are given in Appendix A.

\subsection{Cosmic ray ionization rate and attenuation of CR}

For each of the four cases described above -- X, Y, Z, and Q -- 
we used the spatial variation of the observed intensity in the 1-0 O(2) line 
to constrain the adjustable parameters 
$\zeta_0$, $N_0$ and $\alpha$.  Since our observations primarily probe the 
shielding column density range from $3 - 10 \times 10^{21}\,\rm cm^{-2},$ we adopt a
reference column density of $N_{\rm ref} = 3 \times 10^{21}\,\rm cm^{-2}$ and introduce the parameter
$\zeta_{\rm ref}$, defined as the value of $\zeta_{{\rm H}_2}$ for a shielding column density
of $N_{\rm ref}$:
$$\zeta_{\rm ref} = {\zeta_0 \over (1 + N_{\rm ref}/N_0)^\alpha}$$

Using $\chi^2$ as a measure of goodness of fit, and
adopting flat priors on ${\rm log_{10}}(\zeta_{\rm ref})$, ${\rm log_{10}}(N_0)$,
and $\alpha$ in the intervals $[-17,-13]$, $[19,22]$, and $[0,2]$ respectively, 
we obtained posterior probability distributions 
for the adjustable parameters.  These are represented in Figure Set 8, which also
shows (panel (a)) the best fit to $I_{-7}/N_{22}$ for the strongest CRXH$_2$ line ($v=1-0 O(2)$).
The data points here apply to the CRXH$_2$ component alone, the UVXH$_2$ having been 
subtracted as described in Section 3.

The posterior probability distributions for ${\rm log_{10}} N_0$, $\alpha$, 
and ${\rm log_{10}}(\zeta_{\rm ref})$ are plotted in panels (c), (d) and (e) respectively.
The joint probability distribution for ${\rm log_{10}} N_0$
and ${\rm log_{10}}(\zeta_{\rm ref})$ is show in panel (b), \rd{where red, green and blue
shaded regions indicate the 1, 2, and 3$\,\sigma$ error ``ellipses"}.
Clearly, there is a considerable
degeneracy in ${\rm log_{10}} N_0$
and $\alpha$, as expected since increasing $\alpha$ or decreasing
${\rm log_{10}} N_0$ both increase the degree of attenuation.  Cases 
X, Y, Z and Q appear respectively as Figures 8,  (shown in the printed version), 
8.2, 8.3 and 8.4.  The best-fit parameters are shown by a black dot in panel (b) and are tabulated in Table
3.

\rd{Given the priors discussed above, and the assumption that $\zeta_{{\rm H}_2} (N_s)$ 
is proportional to $(1 + N_{\rm s}/N_0)^{-\alpha},$ we may compute the posterior 
probability distribution for $\zeta_{{\rm H}_2} (N_s)$ as a function of $N_s$.
Results are shown in Figure 9, where the black curve shows 
the median value of $\zeta_{{\rm H}_2}$ (i.e.
the value, $V$, for which the posterior probability of $\zeta_{{\rm H}_2} < V$ equals 0.5).
The dashed and dotted curves show, respectively, the 1 and 2-sigma bounds on 
$\zeta_{{\rm H}_2} (N_s)$ (i.e. the values for which the posterior 
probability of $\zeta_{{\rm H}_2} < V$ is 0.023, 0.16, 0.84, and 0.972 from bottom to top).
As expected, the uncertainties in $\zeta_{{\rm H}_2} (N_s)$ become larger when $N_s$ lies outside
the range that is primarily probed in this study.}
 
\setcounter{figure}{0}
\renewcommand{\thefigure}{8.\arabic{figure}}
\begin{figure}
\centering
\includegraphics[width=0.75\linewidth]{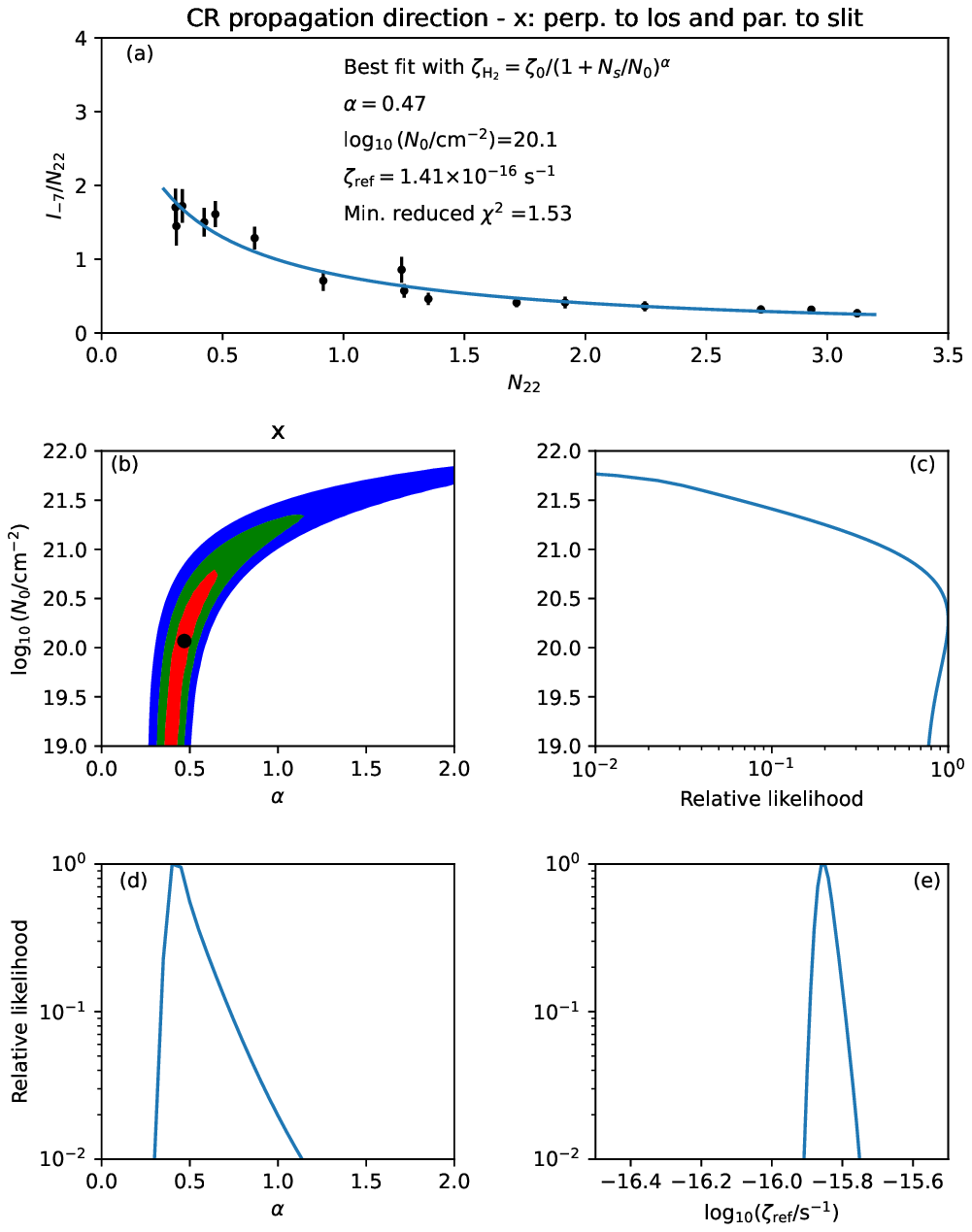}
\caption{Posterior distributions for cosmic-ray attenuation parameters, derived from 
fits to the $v=1$--0 O(2) line. (a) Observed ratio $I_{-7}/N_{22}$ compared to the best-fit model. 
(b) Joint posterior distribution of $\alpha$ and $\log_{10} N_{0}$; best-fit values shown by black dot. 
(c--e) Marginalized distributions for $\log_{10} N_{0}$, $\alpha$, and $\log_{10}\zeta_{\rm ref}$. 
Results apply to case X (propagation direction parallel to the slit).}
\end{figure}

\begin{figure}
\centering
\includegraphics[width=0.85\linewidth]{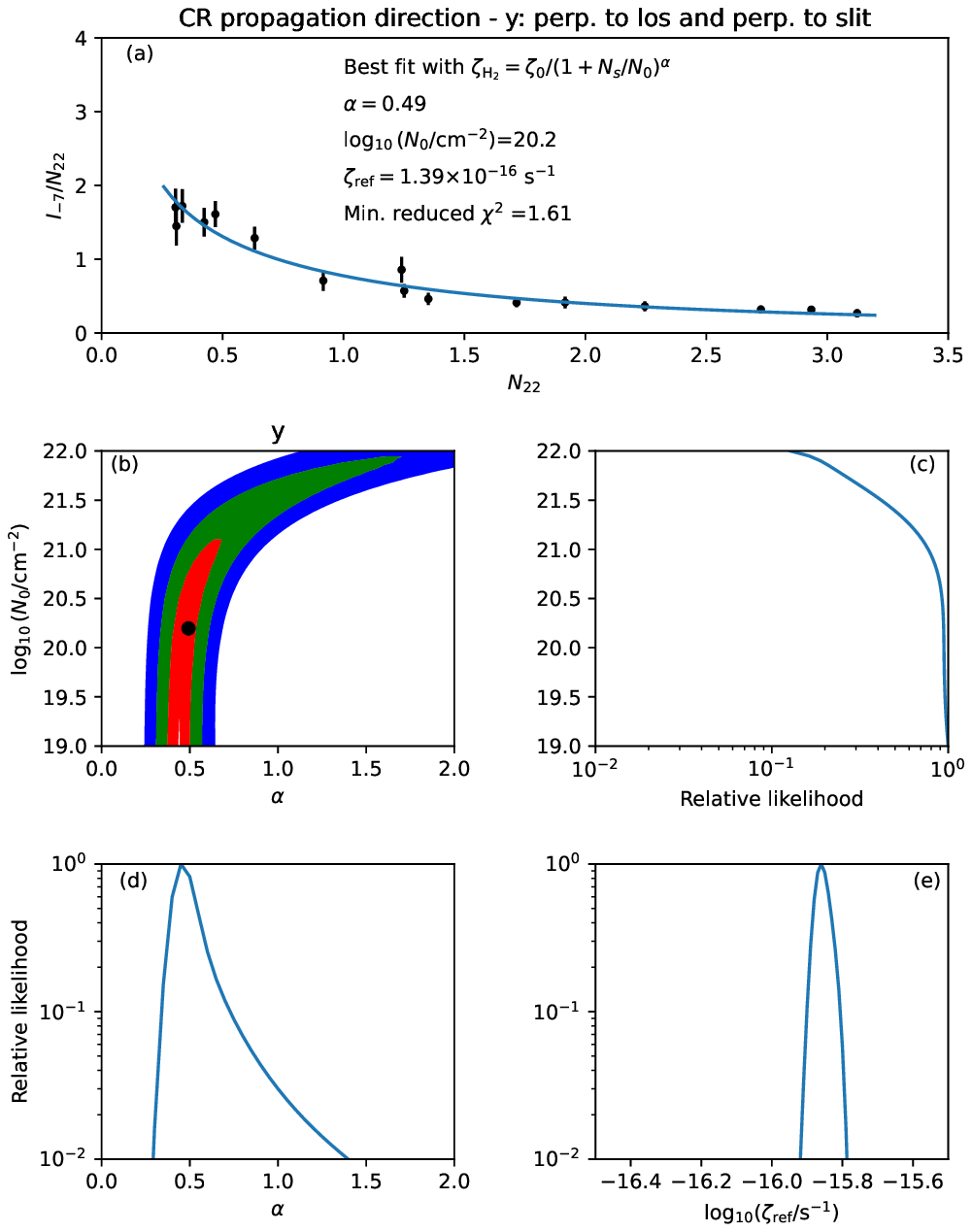}
\caption{Same as Figure 8.1, except for case Y}
\end{figure}

\begin{figure}
\centering
\includegraphics[width=0.85\linewidth]{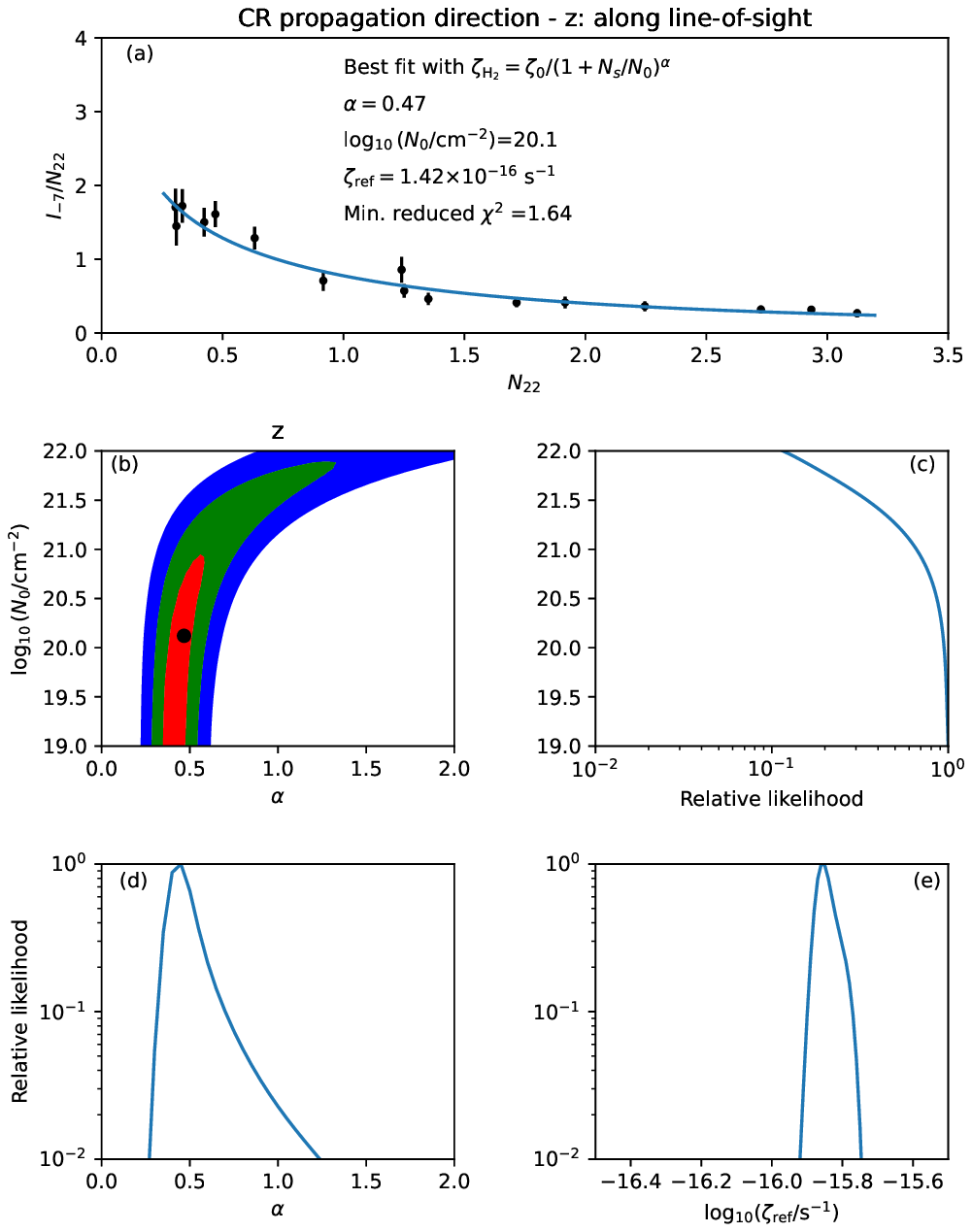}
\caption{Same as Figure 8.1, except for case Z}
\end{figure}

\begin{figure}
\centering
\includegraphics[width=0.85\linewidth]{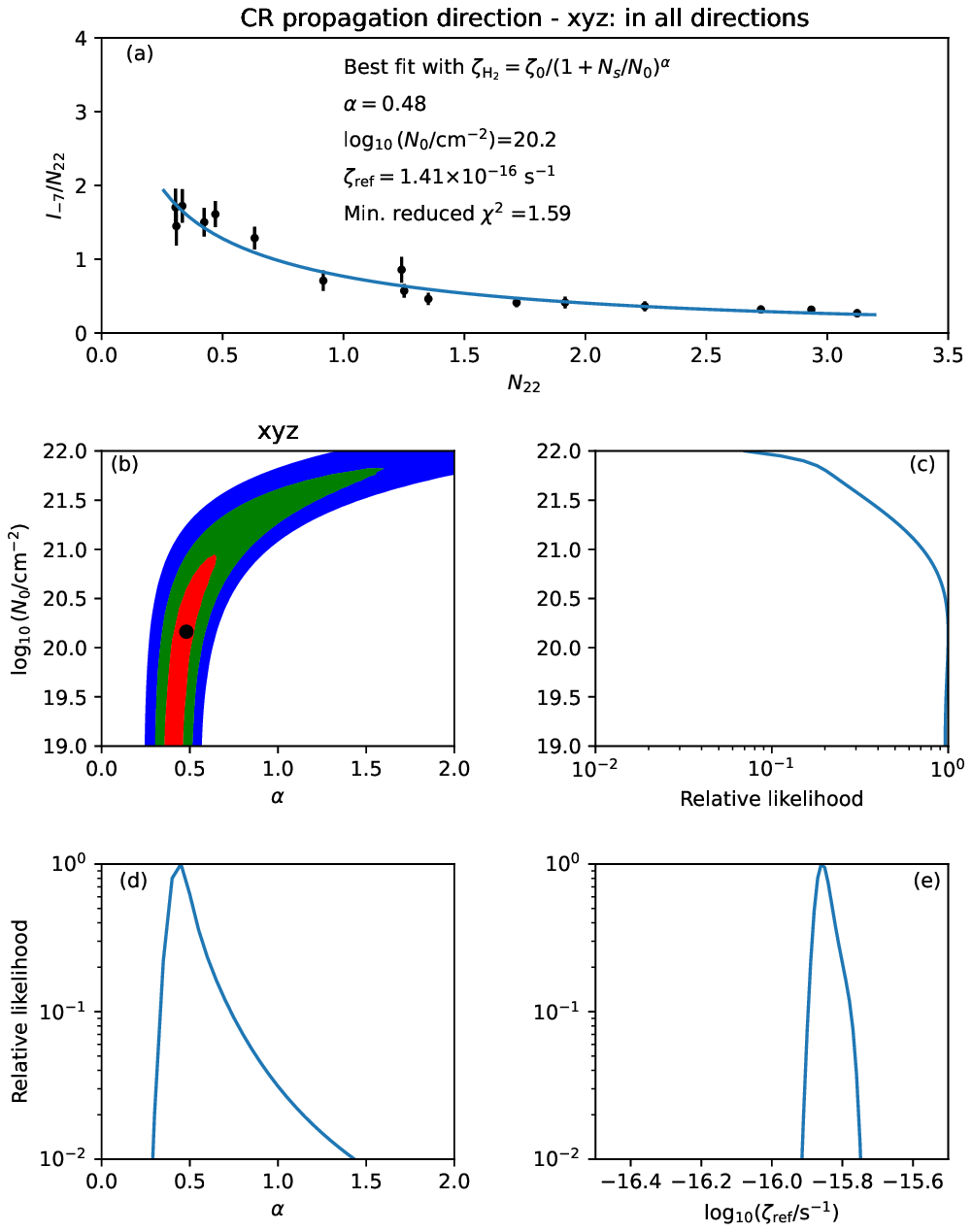}
\caption{Same as Figure 8.1, except for case Q}
\end{figure}
\setcounter{figure}{8}
\renewcommand{\thefigure}{\arabic{figure}}

\begin{figure}
\centering
\includegraphics[width=0.7\linewidth]{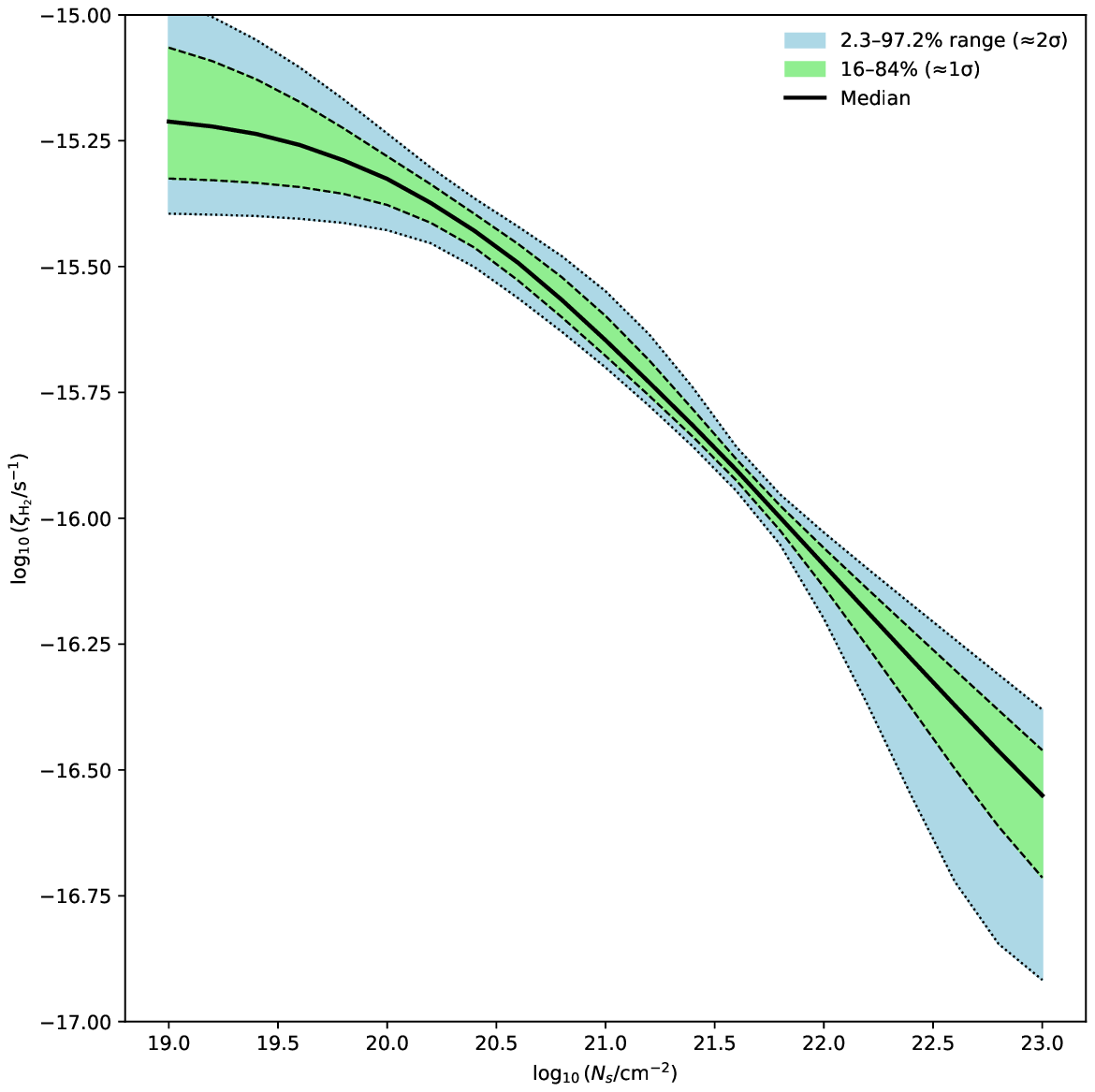}
\caption{\rd{Estimated $\zeta_{{\rm H}_2} (N_s)$ as a function of $N_s$ (see the text for details).}}
\end{figure}

\begin{deluxetable*}{lcccc}
\tablecaption{Best-fit (median) parameters from likelihood analysis\label{tab:bestfit}}
\tablehead{
\colhead{CR propagation direction} &
\colhead{$\alpha$} &
\colhead{$\log_{10}(N_{0}/{\rm cm^{-2}})$} &
\colhead{$\zeta_{\rm H_{2}}$ at $N_{\rm s}=N_{\rm ref}$ (s$^{-1}$)} &
\colhead{Min.\ reduced $\chi^{2}$}
}
\startdata
$x$ (perp.\ to l.o.s., $\parallel$ slit)      & $0.47^{+0.16}_{-0.08}$ & $20.07^{+0.66}_{-0.70}$ & $(1.41^{+0.09}_{-0.07})\times10^{-16}$ & 1.53 \\
$y$ (perp.\ to l.o.s., $\perp$ slit)          & $0.49^{+0.18}_{-0.08}$ & $20.20^{+0.90}_{-0.83}$ & $(1.39^{+0.08}_{-0.07})\times10^{-16}$ & 1.61 \\
$z$ (along line of sight)                      & $0.47^{+0.15}_{-0.08}$ & $20.12^{+0.90}_{-0.77}$ & $(1.42^{+0.12}_{-0.08})\times10^{-16}$ & 1.64 \\
$xyz$ (quasi-isotropic; all directions)        & $0.48^{+0.20}_{-0.08}$ & $20.16^{+0.86}_{-0.79}$ & $(1.41^{+0.12}_{-0.07})\times10^{-16}$ & 1.59 \\
\enddata
\tablecomments{Parameters are derived from fits to the spatial trend of $I_{-7}/N_{22}$ 
in B68 using the attenuation law $\zeta_{\rm H_{2}}=\zeta_{0}\,[1+N_{\rm s}({\rm H_{2}})/N_{0}]^{-\alpha}$. 
Quoted $\zeta_{\rm H_{2}}$ values are evaluated at the reference shielding column $N_{\rm s}({\rm H_{2}})=N_{\rm ref}=3\times10^{21}$~cm$^{-2}$.
The values given are the median values with 68.3\% confidence limits.}
\end{deluxetable*}

\section{Discussion}

\subsection{Attenuation of CR}
Figure Set 8 shows clearly that CR are attenuated as they enter the interior of Barnard 68.
The spatial variation of the line intensities rules out a model ($\alpha=0$) where
$\zeta_{{\rm H}_2}$ is constant and the decline of $I_{-7}/N_{22}$ with $N_{22}$ is the
result of dust extinction alone.   Because of the degeneracy between $\alpha$ and
$N_0$, the probability distribution for $\alpha$ is quite broad, but values smaller than 
\rd{0.31} are robustly excluded ($P \le 1 \times 10^{-2}$) in all four cases (X, Y, Z, and Q.)  

\bl{Although the best-fit parameters we obtain are quite similar in all four geometries,
we can identify case X (propagation direction along the slit) 
as the most { accurate representation of our observations of B68}, given
observational estimates of the magnetic field direction and the assumption that CR propagate
along field lines.  These estimates obtained by Kandori et
al.\ (2020) from near-IR polarimetry, which indicated that the magnetic field direction,
projected onto the plane-of-the-sky, was at position angle $47^\circ \pm 5^\circ$ East of North 
and therefore close to our slit position angle of $57^\circ$.  \ree{Kandori et
al.\ (2020) were also able derive an estimated 
inclination angle of $70^\circ \pm 10^\circ$ to the line-of-sight (i.e.\ concluded
the field is mostly in 
the plane-of-the-sky).  This latter estimate is more model-dependent than the position
angle estimate, because it assumes a specific magnetic field configuration for the source.}}

The attenuation of the CRIR with the column density that we infer
provides important insights into the possible mechanisms of CR transport in 
dense molecular clouds, and may also allow us to put constraints on the energy spectrum of 
interstellar CRs. Here, we summarize and briefly discuss major signatures expected for different transport regimes.

If the transport of CRs penetrating the core were in the regime of free streaming 
(Padovani, Galli \& Glassgold 2009; Padovani et al.\ 2018, hereafter P18), 
their attenuation would depend critically on the slope of the low-energy (non-relativistic) 
spectrum of interstellar CR protons. Assuming a CR proton energy spectrum of the form 
$j(E) \propto E^a/(E_0+E)^b$, with $E_0 = 650\, \rm MeV$ and $b=a+2.7$ (P18), 
the attenuation is predicted to be practically negligible for sufficiently hard spectra 
with $a\simgt -0.2$.  For smaller $a$ the CR attenuation index $\alpha$ decreases
approximately linearly with the spectral index, $a$ \citep[see equation~(33) in][]{Silsbee2019}. 
However, such spectrum would yield a substantially higher 
reference value of the CRIR than the derived $\zeta_{\rm ref}$, as depicted in the 
bottom panel of Figure~10.  
At the same time, the model spectrum 
$\mathcal{L}$ (representing the Voyager data) would substantially underpredict the 
reference value and lead to practically no attenuation of the CRIR. 
The black curves plotted in Figure 9 assume the free-streaming regime, apply specifically
for an assumed value of $650\, \rm MeV$ for $E_0$, and include only the effects of
ionization by CR protons\footnote{While it was shown by P18 that primary electrons 
may substantially enhance the ionization rate at low column densities, this was 
predicated upon the assumption that the energy spectrum of electrons measured by Voyager 
may be extrapolated in energy space with a constant power-law slope to arbitrarily low energy.  
There are two reasons to believe that this overestimates the role of primary electrons.  
First, models of CR acceleration produce a power-law in momentum space, not energy space.  
Second, energy losses due to interaction with the interstellar medium are likely to 
further suppress the flux of CR electrons at low energies.  The possible effect of CR 
electrons on the variation of the ionization rate with the shielding column will be 
investigated elsewhere.}.  They assume that the total ionization rate is enhanced by a 
factor 1.51 due to heavy elements and a factor 1.67 due to secondary ionizations.


Diffusive propagation of CRs, associated with preexisting MHD turbulence that 
is able to resonantly scatter CR particles penetrating molecular clouds, generally 
predicts a steeper attenuation of the CRIR for a given slope of the interstellar 
spectrum \citep{Silsbee2019}. While this feature makes the diffusive transport scenario
more favorable -- in the sense that the data plotted in Figure 9 can be fit better
without any adjustment to $E_0$ --
the very possibility of efficient scattering of non-relativistic 
particles by preexisting turbulence in dense cores remains an open question. The turbulent 
cascade must operate down to extremely small resonant scales, corresponding to the 
CR gyroradius ($\sim0.1$~a.u.\ for a 1~GeV proton), i.e., much beyond the ambipolar 
diffusion scale at which Alfven waves are expected to be 
damped \citep[see, e.g.,][]{Kulsrud1969, Mouschovias2011}. 
Furthermore, the resonant MHD turbulence is carried by ions, whose density in dense 
cores can be dramatically reduced compared to diffuse envelopes (where the ionization 
fraction is set by photoionization); as a result, the diffusive regime of CR propagation 
may cease to operate at column densities over a few 
times $10^{21}$~cm$^{-2}$ \citep[see Figure~1 in][]{Silsbee2019}.\

While streaming into the clouds, interstellar CRs themselves can 
resonantly excite MHD waves in diffuse gas surrounding dense molecular clouds 
\citep{Skilling1976, Ivlev2018}. This phenomenon can lead to self-modulation of 
penetrating CRs: particles are efficiently scattered at the self-generated waves and, 
as a result, the CR spectrum in the cloud is reduced compared to the interstellar 
spectrum at lower energies. Recently, this propagation mechanism was tested with 
GeV gamma-ray observations of nearby giant molecular clouds, showing excellent 
quantitative agreement with the theory \citep{Chernyshov2024}. 
For non-relativistic CRs, the self-modulation is predicted to dramatically reduce 
flux of penetrating particles starting from  column densities of a few times 
$10^{21}$~cm$^{-2}$, which may lead to a very steep attenuation of CRIR. 

In a separate paper \citep{Makarenko_in_prep} we will present a detailed comparative 
analysis of different propagation models for Barnard 68. 

\begin{figure}
\centering
\includegraphics[width=0.9\linewidth]{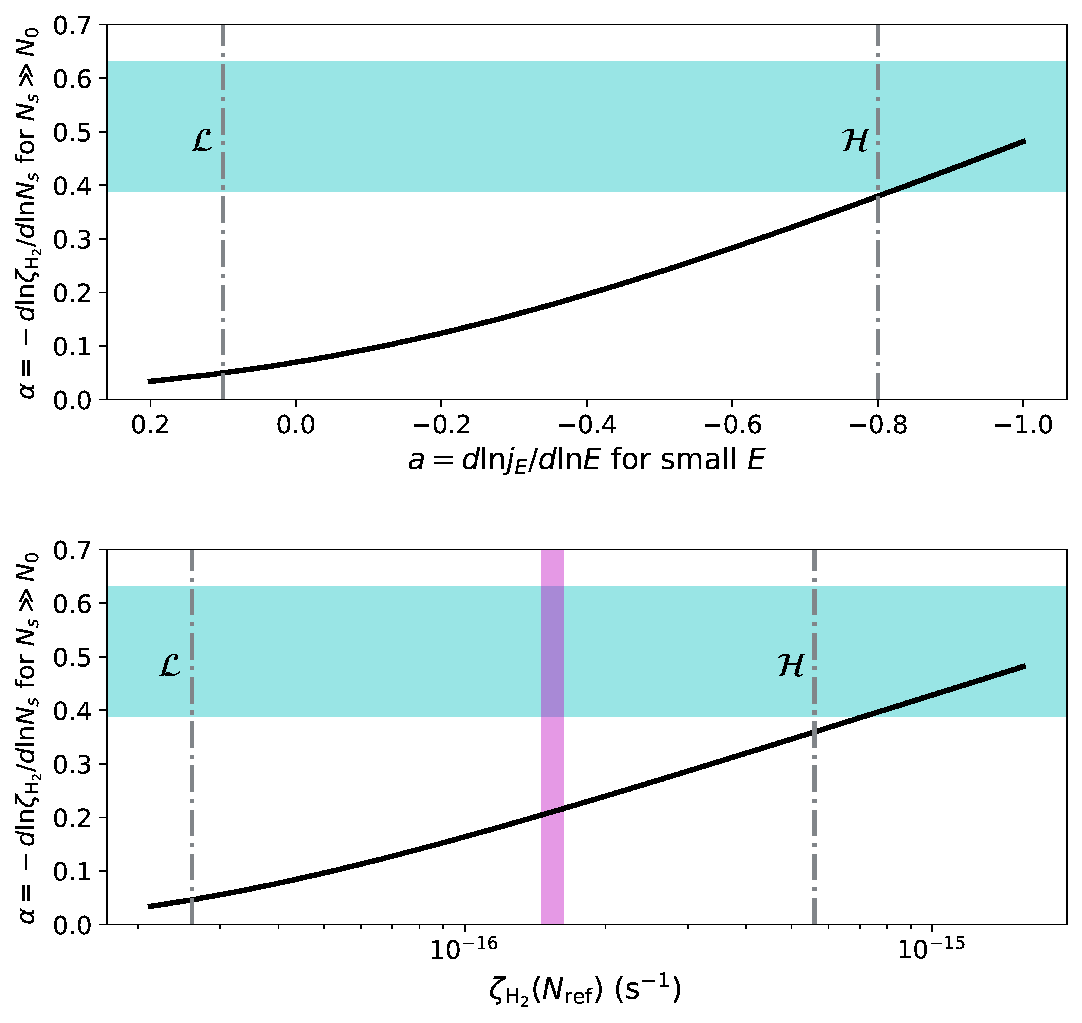}
\caption{Comparison of observed cosmic-ray attenuation with theoretical predictions
\rd{for the free-streaming regime}.
{Top panel:} Dependence of the attenuation index, $\alpha$, on the spectral slope,
$a$, of the low-energy cosmic-ray proton spectrum. 
The horizontal shaded band indicates the
68\% confidence interval for $\alpha$ inferred from our fits to B68.
{Bottom panel:} Predicted cosmic-ray ionization rate at the reference column
density, $\zeta_{\rm ref}$, for representative model spectra \citep{Padovani2018},
compared to the value inferred for B68 (vertical shaded band).
}
\end{figure}

\subsection{Cosmic-ray ionization rate}

\bl{As indicated in Table 3, the CRIR at the reference
shielding column of $3 \times 10^{21} \, \rm cm^{-2}$ is $\zeta_{\rm H_2} \sim 1.4 \times 10^{-16}\rm \,s^{-1}$, 
including both primary and secondary ionizations.  This value, which is only
very weakly dependent on the assumed CR propagation direction, is somewhat higher
than the values typically obtained from recent analyses 
of the H$_3^+$ abundances in diffuse and translucent molecular clouds.   
Taken together, the studies of Obolentseva et al.\ (2024) and Indriolo et al.\ (2025) 
have identified 16 sight-lines that exhibit H$_3^+$ absorption and yield $\zeta_{{\rm H}_2}$ 
values ranging from $2.7 - 11.5 \times 10^{-17}\, \rm s^{-1}$.  The average value in these 16 clouds
was  $5.3 \times 10^{-17}\, \rm s^{-1}$, a factor of 3 smaller than the value determined for B68
\rd{(although the largest value determined from H$_3^+$ absorption observations
was smaller by only $18\%$)}.  { Properly accounting for cosmic ray attenuation in the 
H$_3^+$ method would probably enhance the discrepancy somewhat since the typical shielding column 
densities probed in Obolentseva (2024) and Indriolo (2025) are a few times lower than our 
reference shielding column.}
The two methods being compared here, the CRXH$_2$ method
and the H$_3^+$ method, are entirely independent and undoubtedly have different 
systematics.  In particular, the CRIR determination from H$_3^+$ is proportional to the
assumed volume density in the absorbing cloud, the latter being derived from 
extinction maps. { If density estimates are taken from measurements of C$_2$ absorption 
line ratios instead of from extinction maps, the CRIR inferred from H$_3^+$ increases
systematically by a typical factor of 2 -- 3 (Indriolo et al.\ (2025), 
bringing the range inferred from H$_3^+$ into comfortable agreement with our estimate for B68.}
In any case, it remains unclear to what extent any differences between the 
CRIR derived in B68 and those typically obtained from H$_3^+$ observations  
reflect a true variation in the CRIR and to what extent it is a systematic discrepancy
between the different methods.  
Further observations using the CRXH$_2$ method are
urgently needed to extend the analysis beyond the single source considered in the present
study.}

\subsection{Para-H$_2$ line ratios}

{ Our observations of multiple para-H$_2$ lines provide a valuable probe of
the excitation mechanism.  In particular, the $v=2-1$ O(2) / $v=1-0$ O(2) intensity ratio, 
which we will refer to as $R_{21}$,
is a powerful discriminant between CR- and UV-excitation.  For the difference spectrum, B68 - OFF,
the intensities given in Table 2 above yield $R_{21} = 0.197 \pm 0.029$.  This value lies
between the predictions for UVXH$_2$ and CRXH$_2$, which differ significantly because 
of the different physical mechanisms involved: UV radiation pumps electronic states, 
leading to a radiative cascade, whereas CRXH$_2$ emissions are excited directly by CR secondary electrons 
that excite to $v=1$ much more efficiently than to $v=2$.  UVXH$_2$ emissions 
yield $R_{21} = 0.58$ (S88),
a value that we confirmed with the Meudon PDR code (Le Petit et al.\ 2006) and found
to vary by less than 3$\%$ over many orders of magnitude in the assumed
gas pressure.  Because $R_{21}$ involves lines 
of para-H$_2$ alone, the UVXH$_2$ prediction is independent of the OPR in the UV-excited
gas.  For CRXH$_2$, the expected ratio is $R_{21} =0.038$, based on cross-sections given by
Zammit et al.\ (2017) and Scarlett et al.\ (2023). 
Thus the value of $R_{21}$ we observed suggests 
strongly that both CRXH$_2$ and UVXH$_2$ contribute to the signal
measured in the difference spectrum.
The pure UVXH$_2$ prediction is ruled out at the 13$\sigma$ level, providing compelling
evidence for the presence of a CRXH$_2$ component.}

\subsection{Approximations made in the analysis}

\bl{We complete the discussion section with a review of several approximations made in the analysis.}
\vskip 0.1 true in
\begin{enumerate}  
\item We have assumed the B68 cloud to be spherically-symmetric.  While the projected 
extinction map shows only modest asymmetries, the structure along the line-of-sight is unknown.

\item We have adopted specific assumptions about the properties of the dust.  In 
particular, we used the KP5 extinction curve tabulated in 
Pontoppidan et al.\ (2024), which includes the effects of ice absorption features, and
assumed an $N_{22}/A_V$ ratio of 0.094.  

\item In computing the attenuation of the CRXH$_2$ emissions, we applied a simple
extinction correction: this treatment neglects the contribution of scattered radiation to the
observed flux.

\item We assumed the gas within B68 to be fully molecular.

\item We assumed that CRXH$_2$ emission makes a negligible contribution to the 
ortho-H$_2$ line intensities we observed, { and that the relative line strengths
of the H$_2$ emissions attributable to the
the UVXH$_2$ component in B68 are the same as those in the background region.}

\noindent{ In a future study, we will use detailed PDR models and dust radiative transfer models
with scattering to remove the need for approximations 3, 4 and 5.}

\end{enumerate}

\section{Summary}  

Our JWST/NIRSpec observations of the starless dark cloud Barnard~68 have revealed new insights into the excitation of molecular hydrogen by cosmic rays. The main results of this study are as follows:  

\begin{enumerate}  
\item Multiple rovibrational H$_2$ lines attributable to cosmic-ray excitation were detected across B68. 
Their spectra, dominated by para-H$_2$ transitions, clearly distinguish them from UV-pumped fluorescence.  

\item By comparing the B68 spectra to a nearby OFF position, where the relative line strengths are in excellent agreement with models for UV excitation, 
we obtained line intensities that are robustly attributable to CR excitation alone.  

\item The CRXH$_2$ line intensities vary systematically with position across the cloud, declining relative to the column density with increasing depth. 
This trend cannot be explained by dust extinction alone, and demonstrates the attenuation of cosmic rays within B68.  

\item A Bonnor--Ebert sphere model was used to reproduce the observed spatial dependence of the CRXH$_2$ lines. 
Fits to the data require attenuation of the cosmic-ray ionization rate with depth, with values of the attenuation parameter $\alpha < \rd{0.31}$ strongly excluded.  

\item At a reference column density of $N({\rm H}_2) = 3 \times 10^{21}\,{\rm cm}^{-2}$, 
the best-fit models yield a cosmic-ray ionization rate of $\zeta_{\rm H_2} \sim 1.4 \times 10^{-16}\,{\rm s}^{-1}$, 
a factor of $\approx 3$ higher than the average values derived from H$_3^+$ absorption in diffuse and translucent clouds.  


\item These results establish CR-excited H$_2$ emission as a 
powerful new diagnostic of cosmic-ray penetration in dense molecular gas, with important 
consequences for our understanding of ionization, chemistry, and the role of low-energy cosmic rays in star-forming regions.  
\end{enumerate}  

\acknowledgments{This work is based on observations made with the
NASA/ESA/CSA James Webb Space Telescope. The
data were obtained from the Mikulski Archive for Space
Telescopes at the Space Telescope Science Institute,
which is operated by the Association of Universities
for Research in Astronomy, Inc., under NASA contract
NAS 5-03127 for JWST. 
These observations are associated 
with program \#5064. We are grateful to J.\ Alves for providing
an unpublished extinction map of B68 in digital format.
Support for D.A.N. and K.S.
was provided by NASA through a grant from the Space
Telescope Science Institute.  MP acknowledges the INAF grant { 2023 
MERCATOR (``MultiwavelEngth signatuRes of Cosmic rAys in sTar-fOrming Regions'')
and} the INAF grant 2024 ENERGIA 
(``ExploriNg low-Energy cosmic Rays throuGh theoretical InvestigAtions at INAF'').
SB acknowledges financial support from the Israeli Science Foundation grant 2071540, and 
the German Israeli Science Foundation Grant 1568, and the Technion.
BALG is supported by the German Research Foundation (DFG) in the form of an 
Emmy Noether Research Group - DFG project \#542802847 (GA 3170/3-1).
{ CHR acknowledges the support of the Deutsche Forschungsgemeinschaft 
(DFG, German Research Foundation) Research Unit “Transition discs” - 325594231. 
CHR is grateful for support from the Max Planck Society.}}

\software{JWST Calibration Pipeline (v1.16.1) 
with Calibration Reference Data System context 1303 (Bushouse et al.\ 2024), astropy (Astropy collaboration 2013, 2018, 2022), 
{scipy},
NASA Astrophysics Data System (ADS), OpenAI GPT-5.} 

\section*{Data Availability}
The data underlying this article are publicly available from the Mikulski Archive for 
Space Telescopes (MAST) at the Space Telescope Science Institute. These observations 
were obtained as part of JWST program 5064 (PI: Bialy). { The specific observations analyzed 
can be accessed via \dataset[doi: 10.17909/1bw2-ga13]{https://doi.org/10.17909/1bw2-ga13}}.

\appendix
\section{Expressions for $\zeta_{{\rm H}_2}$ }

The expressions for $\zeta_{{\rm H}_2}$ in the four cases discussed in Section 4 are as follows

\begin{itemize}
\item Case X: CR propagation in the plane-of-the-sky 
along the slit direction, with an equal contribution in the $-x$ and $+x$
directions.  Here $\zeta_{{\rm H}_2}(x,z) = \zeta_0 [F_s(N_{+x}) + F_s (N_{-x})] / 2,$
with $$N_{+x} = \int_{-x_0}^{x} n_{{\rm H}_2}([x^{\prime 2}+z^2]^{1/2}) dx^\prime$$
$$N_{-x} = \int_{x}^{x_0} n_{{\rm H}_2}([x^{\prime 2}+z^2]^{1/2}) dx^\prime$$
and $x_0=(r_{\rm out}^2-z^2)^{1/2}$.  $F_s$ is the adopted shielding function 
$(1 + N_s/N_0)^{-\alpha}$ (Section 4), and $n_{{\rm H}_2}(r)$ is the H$_2$ particle density 
as a function of the radial coordinate.

\item Case Y: CR propagation in the plane-of-the-sky 
perpendicular to the slit direction, with an equal contribution in the $-y$ and $+y$
directions.  Here $\zeta_{{\rm H}_2}(x,z) = \zeta_0 [F_s(N_{+y}) + F_s(N_{-y})] / 2,$
with $$\rd{N_{+y} = N_{-y} = \int_{0}^{y_0} n_{{\rm H}_2}([x^2+y^{\prime 2}+z^2]^{1/2}) dy^\prime}$$
and $y_0=(r_{\rm out}^2-x^2-z^2)^{1/2}$

\item Case Z: CR propagation along the line-of-sight, with an equal contribution in the $-z$ and $+z$
directions.  Here $\zeta_{{\rm H}_2}(x,z) = \zeta_0 [F_s(N_{+z}) + F_s(N_{-z})] / 2,$
with $$N_{+z} = \int_{-z_0}^{z} n_{{\rm H}_2}([x^2+z^{\prime 2}]^{1/2}) dz^\prime$$
$$N_{-z} = \int_{z}^{z_0} n_{{\rm H}_2}([x^2+z^{\prime 2}]^{1/2}) dz^\prime$$
\rd{and $z_0=(r_{\rm out}^2-x^2)^{1/2}$}.

\item Case Q: Quasi-isotropic CR propagation, with an equal contribution in the 
$\pm x$, $\pm y$, and $\pm z$
directions.  
Here $\zeta_{{\rm H}_2}(x,z) = \zeta_0 [F_s(N_{+x}) + F_s(N_{-x}) + F_s(N_{+y}) + F_s(N_{-y}) 
+ F_s(N_{+z}) + F_s(N_{-z})] / 6$

\end{itemize}


\begin{thebibliography}{}

\bibitem[Alves et al.(2001)]{2001Natur.409..159A} 
Alves, J.~F., Lada, C.~J., \& Lada, E.~A.\ 2001, \nat, 409, 6817, 159, doi:10.1038/35051509

\bibitem[Astropy Collaboration et al.(2013)]{2013A&A...558A..33A} Astropy Collaboration, Robitaille, T.~P., Tollerud, E.~J., et al.\ 2013, \aap, 558, A33. doi:10.1051/0004-6361/201322068

\bibitem[Astropy Collaboration et al.(2018)]{2018AJ....156..123A} Astropy Collaboration, Price-Whelan, A.~M., Sip{\H{o}}cz, B.~M., et al.\ 2018, \aj, 156, 3, 123. doi:10.3847/1538-3881/aabc4f

\bibitem[Astropy Collaboration et al.(2022)]{2022ApJ...935..167A} Astropy Collaboration, Price-Whelan, A.~M., Lim, P.~L., et al.\ 2022, \apj, 935, 2, 167. doi:10.3847/1538-4357/ac7c74

\bibitem[Bialy(2020)]{2020CmPhy...3...32B} 
Bialy, S.\ 2020, Communications Physics, 3, 1, 32, doi:10.1038/s42005-020-0293-7

\bibitem[Bialy et al.(2025)]{2025ApJ...tbd..tbdB} 
Bialy, S., et al.\ 2025, Nature Astronomy, submitted (arXiv:2508.20168v1) 

\bibitem[Bohlin et al.(1978)]{1978ApJ...224..132B} 
Bohlin, R.~C., Savage, B.~D., \& Drake, J.~F.\ 1978, \apj, 224, 132, doi:10.1086/156357

\bibitem[Bushouse et al.(2024)]{bushouse2024jwstpipe}
Bushouse, H., Eisenhamer, J., Dencheva, N., et al. 2024,
\textit{JWST Calibration Pipeline},
Zenodo, \href{https://doi.org/10.5281/zenodo.14153298}{doi:10.5281/zenodo.14153298}

\bibitem[Chernyshov et al.(2024)]{Chernyshov2024} 
Chernyshov, D.~O., Ivlev, A.~V., \& Kiselev, A.~M.\ 2024, \prd, 110, 043012, doi:10.1103/PhysRevD.110.043012

\bibitem[Draine(2003)]{2003ApJ...598.1017D} 
Draine, B.~T.\ 2003, \apj, 598, 2, 1017, doi:10.1086/379118

\bibitem[Evans et al.(2009)]{2009ApJS..181..321E} 
Evans, N.~J., Dunham, M.~M., J{\o}rgensen, J.~K., et al.\ 2009, \apjs, 181, 2, 321. doi:10.1088/0067-0049/181/2/321

\bibitem[Hotzel et al.(2002)]{Hotzel2002}
Hotzel, S., Harju, J., Juvela, M., Mattila, K., \& Haikala, L. K. 2002, \aap, 391, 275,
doi:10.1051/0004-6361:20020877

\bibitem[Indriolo et al.(2025)]{Indriolo2025}
Indriolo, N., Obolentseva, M., Ivlev, A. V., Silsbee, K., Neufeld, D. A., Caselli, P.,
Edenhofer, G., Bisbas, T. G., \& Lomeli, D. 2025, \apj, submitted

\bibitem[Ivlev et al.(2015)]{Ivlev2015}
Ivlev, A. V., Padovani, M., Galli, D., \& Caselli, P. 2015, \apj, 812, 135,
doi:10.1088/0004-637X/812/2/135

\bibitem[Ivlev et al.(2018)]{Ivlev2018} 
Ivlev, A.~V., Dogiel, V.~A., Chernyshov, D.~O., Caselli, P., Ko, C.-M., \& Cheng, K.~S.\ 2018, \apj, 855, 23, doi:10.3847/1538-4357/aaadb9

\bibitem[Jakobsen et al.(2022)]{2022A&A...661A..80J} 
Jakobsen, P., Ferruit, P., Alves de Oliveira, C., et al.\ 2022, \aap, 661, A80, doi:10.1051/0004-6361/202142663

\bibitem[Kandori et al.(2020)]{2020PASJ...72....8K} 
Kandori, R., Tamura, M., Saito, M., et al.\ 2020, \pasj, 72, 1, 8. doi:10.1093/pasj/psz127

\bibitem[Kulsrud \& Pearce(1969)]{Kulsrud1969} 
Kulsrud, R., \& Pearce, W.~P.\ 1969, \apj, 156, 445, doi:10.1086/149981

\bibitem[Lacy et al.(2017)]{2017ApJ...838...66L} 
Lacy, J.~H., Sneden, C., Kim, H., et al.\ 2017, \apj, 838, 1, 66, doi:10.3847/1538-4357/aa6247

\bibitem[Le Petit et al.(2006)]{2006ApJS..164..506L} Le Petit, F., Nehm{\'e}, C., Le Bourlot, J., et al.\ 2006, \apjs, 164, 2, 506. doi:10.1086/503252

\bibitem[Makarenko et al.(2025)]{Makarenko_in_prep} 
Makarenko, E., et al.\ 2025, in preparation

\bibitem[Mouschovias et al.(2011)]{Mouschovias2011} 
Mouschovias, T.~Ch., Ciolek, G.~E., \& Morton, S.~A.\ 2011, \mnras, 415, 1751, doi:10.1111/j.1365-2966.2011.18817.x

\bibitem[Nielbock et al.(2012)]{2012A&A...547A..11N} Nielbock, M., Launhardt, R., Steinacker, J., et al.\ 2012, \aap, 547, A11. doi:10.1051/0004-6361/201219139

\bibitem[Obolentseva et al.(2024)]{Obolentseva2024} 
Obolentseva, M., Ivlev, A.~V., Silsbee, K., Neufeld, D.~A., Caselli, P., Edenhofer, G., Indriolo, N., Bisbas, T.~G., \& Lomeli, D.\ 2024, \apj, 973, 142, doi:10.3847/1538-4357/ad71ce

\bibitem[Padovani et al.(2009)]{Padovani2009}
Padovani, M., Galli, D., \& Glassgold, A. E. 2009, \aap, 501, 619,
doi:10.1051/0004-6361/200911794

\bibitem[Padovani et al.(2018)]{Padovani2018} 
Padovani, M., Ivlev, A.~V., Galli, D., \& Caselli, P.\ 2018, \aap, 614, A111, doi:10.1051/0004-6361/201732202

\bibitem[Padovani et al.(2022)]{2022A&A...658A.189P} 
Padovani, M., Bialy, S., Galli, D., et al.\ 2022, \aap, 658, A189, doi:10.1051/0004-6361/202142560

\bibitem[Pontoppidan et al.(2024)]{2024RNAAS...8...68P} 
Pontoppidan, K.~M., Evans, N., Bergner, J., et al.\ 2024, RNAAS, 8, 3, 68, doi:10.3847/2515-5172/ad303f

\bibitem[Roy et al.(2014)]{2014A&A...562A.138R} Roy, A., Andr{\'e}, P., Palmeirim, P., et al.\ 2014, \aap, 562, A138. doi:10.1051/0004-6361/201322236

\bibitem[Scarlett et al.(2023)]{2023PhRvA.107f2804S} 
Scarlett, L.~H., Rehill, U.~S., Zammit, M.~C., et al.\ 2023, \pra, 107, 6, 062804. doi:10.1103/PhysRevA.107.062804

\bibitem[Skilling \& Strong(1976)]{Skilling1976} 
Skilling, J., \& Strong, A.~W.\ 1976, \aap, 53, 253.

\bibitem[Silsbee \& Ivlev(2019)]{Silsbee2019} 
Silsbee, K., \& Ivlev, A.~V.\ 2019, \apj, 879, 14, doi:10.3847/1538-4357/ab22b4

\bibitem[Storey \& Hummer(1995)]{1995MNRAS.272...41S} 
Storey, P.~J. \& Hummer, D.~G.\ 1995, \mnras, 272, 1, 41, doi:10.1093/mnras/272.1.41

\bibitem[Sternberg(1988)]{1988ApJ...332..400S} 
Sternberg, A.\ 1988, \apj, 332, 400, doi:10.1086/166664

\bibitem[Zammit et al.(2017)]{2017PhRvA..95b2708Z} 
Zammit, M.~C., Savage, J.~S., Fursa, D.~V., et al.\ 2017, \pra, 95, 2, 022708, doi:10.1103/PhysRevA.95.022708

\bibitem[Zhu et al.(2017)]{2017MNRAS.471.3494Z} 
Zhu, H., Tian, W., Li, A., et al.\ 2017, \mnras, 471, 3, 3494, doi:10.1093/mnras/stx1580

\end{thebibliography}
\end{document}